\documentclass[a4paper,11pt]{article}

\makeatletter

\gdef\@fpheader{\newline}
\gdef\@journal{jhep}

\RequirePackage{amsmath}
\RequirePackage{amssymb}
\RequirePackage{epsfig}
\RequirePackage{graphicx}
\RequirePackage[numbers,sort&compress]{natbib}
\RequirePackage{color}
\RequirePackage[dvipdfmx
,colorlinks=true
,urlcolor=blue
,anchorcolor=blue
,citecolor=blue
,filecolor=blue
,linkcolor=blue
,menucolor=blue
,pagecolor=blue
,linktocpage=true
,pdfproducer=medialab
,pdfa=true
]{hyperref}

\newif\ifnotoc\notocfalse
\newif\ifemailadd\emailaddfalse
\newif\iftoccontinuous\toccontinuousfalse

\def\@subheader{\@empty}
\def\@keywords{\@empty}
\def\@abstract{\@empty}
\def\@xtum{\@empty}
\def\@dedicated{\@empty}
\def\@arxivnumber{\@empty}
\def\@collaboration{\@empty}
\def\@collaborationImg{\@empty}
\def\@proceeding{\@empty}
\def\@preprint{\@empty}

\newcommand{\subheader}[1]{\gdef\@subheader{#1}}
\newcommand{\keywords}[1]{\if!\@keywords!\gdef\@keywords{#1}\else%
\PackageWarningNoLine{\jname}{Keywords already defined.\MessageBreak Ignoring last definition.}\fi}
\renewcommand{\abstract}[1]{\gdef\@abstract{#1}}
\newcommand{\dedicated}[1]{\gdef\@dedicated{#1}}
\newcommand{\arxivnumber}[1]{\gdef\@arxivnumber{#1}}
\newcommand{\proceeding}[1]{\gdef\@proceeding{#1}}
\newcommand{\xtumfont}[1]{\textsc{#1}}
\newcommand{\correctionref}[3]{\gdef\@xtum{\xtumfont{#1} \href{#2}{#3}}}
\newcommand\jname{JHEP}
\newcommand\acknowledgments{\section*{Acknowledgments}}

\newcommand\preprint[1]{\gdef\@preprint{\hfill #1}}

\newcommand\note[2][]{%
\if!#1!%
\stepcounter{footnote}\footnotetext{#2}%
\else%
{\renewcommand\thefootnote{#1}%
\footnotetext{#2}}%
\fi}

\newtoks\auth@toks
\renewcommand{\author}[2][]{%
  \if!#1!%
    \auth@toks=\expandafter{\the\auth@toks#2\ }%
  \else
    \auth@toks=\expandafter{\the\auth@toks#2$^{#1}$\ }%
  \fi
}

\newtoks\affil@toks\newif\ifaffil\affilfalse
\newcommand{\affiliation}[2][]{%
\affiltrue
  \if!#1!%
    \affil@toks=\expandafter{\the\affil@toks{\item[]#2}}%
  \else
    \affil@toks=\expandafter{\the\affil@toks{\item[$^{#1}$]#2}}%
  \fi
}

\newtoks\email@toks\newcounter{email@counter}%
\newcommand{\emailAdd}[1]{%
\emailaddtrue%
\ifnum\theemail@counter>0\email@toks=\expandafter{\the\email@toks, \@email{#1}}%
\else\email@toks=\expandafter{\the\email@toks\@email{#1}}%
\fi\stepcounter{email@counter}}
\newcommand{\@email}[1]{\href{mailto:#1}{\tt #1}}

\newcommand*\collaboration[1]{\gdef\@collaboration{#1}}
\newcommand*\collaborationImg[2][]{\gdef\@collaborationImg{#2}}

\newcommand\afterLogoSpace{\smallskip}
\newcommand\afterSubheaderSpace{\vskip3pt plus 2pt minus 1pt}
\newcommand\afterProceedingsSpace{\vskip21pt plus0.4fil minus15pt}
\newcommand\afterTitleSpace{\vskip23pt plus0.06fil minus13pt}
\newcommand\afterRuleSpace{\vskip23pt plus0.06fil minus13pt}
\newcommand\afterCollaborationSpace{\vskip3pt plus 2pt minus 1pt}
\newcommand\afterCollaborationImgSpace{\vskip3pt plus 2pt minus 1pt}
\newcommand\afterAuthorSpace{\vskip5pt plus4pt minus4pt}
\newcommand\afterAffiliationSpace{\vskip3pt plus3pt}
\newcommand\afterEmailSpace{\vskip16pt plus9pt minus10pt\filbreak}
\newcommand\afterXtumSpace{\par\bigskip}
\newcommand\afterAbstractSpace{\vskip16pt plus9pt minus13pt}
\newcommand\afterKeywordsSpace{\vskip16pt plus9pt minus13pt}
\newcommand\afterArxivSpace{\vskip3pt plus0.01fil minus10pt}
\newcommand\afterDedicatedSpace{\vskip0pt plus0.01fil}
\newcommand\afterTocSpace{\bigskip\medskip}
\newcommand\afterTocRuleSpace{\bigskip\bigskip}
\newlength{\affiliationsSep}
\newcommand\beforetochook{\pagestyle{myplain}\pagenumbering{roman}}

\DeclareFixedFont\trfont{OT1}{phv}{b}{sc}{11}

\renewcommand\maketitle{
\pagestyle{empty}
\thispagestyle{titlepage}
\setcounter{page}{0}
\noindent{\small\scshape\@fpheader}\@preprint\par
\afterLogoSpace
\if!\@subheader!\else\noindent{\trfont{\@subheader}}\fi
\afterSubheaderSpace
\if!\@proceeding!\else\noindent{\sc\@proceeding}\fi
\afterProceedingsSpace
{\LARGE\flushleft\sffamily\bfseries\@title\par}
\afterTitleSpace
\hrule height 1.5\p@%
\afterRuleSpace
\if!\@collaboration!\else
{\Large\bfseries\sffamily\raggedright\@collaboration}\par
\afterCollaborationSpace
\fi
\if!\@collaborationImg!\else
{\normalsize\bfseries\sffamily\raggedright\@collaborationImg}\par
\afterCollaborationImgSpace
\fi
{\bfseries\raggedright\sffamily\the\auth@toks\par}
\afterAuthorSpace
\ifaffil\begin{list}{}{%
\setlength{\leftmargin}{0.28cm}%
\setlength{\labelsep}{0pt}%
\setlength{\itemsep}{\affiliationsSep}%
\setlength{\topsep}{-\parskip}}
\itshape\small%
\the\affil@toks
\end{list}\fi
\afterAffiliationSpace
\ifemailadd 
\noindent\hspace{0.28cm}\begin{minipage}[l]{.9\textwidth}
\begin{flushleft}
\textit{E-mail:} \the\email@toks
\end{flushleft}
\end{minipage}
\else 
\PackageWarningNoLine{\jname}{E-mails are missing.\MessageBreak Plese use \protect\emailAdd\space macro to provide e-mails.}
\fi
\afterEmailSpace
\if!\@xtum!\else\noindent{\@xtum}\afterXtumSpace\fi
\if!\@abstract!\else\noindent{\renewcommand\baselinestretch{.9}\textsc{Abstract:}}\ \@abstract\afterAbstractSpace\fi
\if!\@keywords!\else\noindent{\textsc{Keywords:}} \@keywords\afterKeywordsSpace\fi
\if!\@arxivnumber!\else\noindent{\textsc{ArXiv ePrint:}} \href{http://arxiv.org/abs/\@arxivnumber}{\@arxivnumber}\afterArxivSpace\fi
\if!\@dedicated!\else\vbox{\small\it\raggedleft\@dedicated}\afterDedicatedSpace\fi
\ifnotoc\else
\iftoccontinuous\else\newpage\fi
\beforetochook\hrule
\tableofcontents
\afterTocSpace
\hrule
\afterTocRuleSpace
\fi
\setcounter{footnote}{0}
\pagestyle{myplain}\pagenumbering{arabic}
} 

\renewcommand{\baselinestretch}{1.1}\normalsize
\divide\@tempdima\baselineskip
\@tempcnta=\@tempdima
\skip\@mpfootins = \skip\footins
\renewcommand{\@dotsep}{10000}

\newcommand\ps@myplain{
\pagenumbering{arabic}
\renewcommand\@oddfoot{\hfill-- \thepage\ --\hfill}
\renewcommand\@oddhead{}}
\let\ps@plain=\ps@myplain

\newcommand\ps@titlepage{\renewcommand\@oddfoot{}\renewcommand\@oddhead{}}

\numberwithin{equation}{section}

\renewcommand\section{\@startsection{section}{1}{\z@}%
                                   {-3.5ex \@plus -1.3ex \@minus -.7ex}%
                                   {2.3ex \@plus.4ex \@minus .4ex}%
                                   {\normalfont\large\bfseries}}
\renewcommand\subsection{\@startsection{subsection}{2}{\z@}%
                                   {-2.3ex\@plus -1ex \@minus -.5ex}%
                                   {1.2ex \@plus .3ex \@minus .3ex}%
                                   {\normalfont\normalsize\bfseries}}
\renewcommand\subsubsection{\@startsection{subsubsection}{3}{\z@}%
                                   {-2.3ex\@plus -1ex \@minus -.5ex}%
                                   {1ex \@plus .2ex \@minus .2ex}%
                                   {\normalfont\normalsize\bfseries}}
\renewcommand\paragraph{\@startsection{paragraph}{4}{\z@}%
                                   {1.75ex \@plus1ex \@minus.2ex}%
                                   {-1em}%
                                   {\normalfont\normalsize\bfseries}}
\renewcommand\subparagraph{\@startsection{subparagraph}{5}{\parindent}%
                                   {1.75ex \@plus1ex \@minus .2ex}%
                                   {-1em}%
                                   {\normalfont\normalsize\bfseries}}

\def\fnum@figure{\textbf{\figurename\nobreakspace\thefigure}}
\def\fnum@table{\textbf{\tablename\nobreakspace\thetable}}

\long\def\@makecaption#1#2{%
  \vskip\abovecaptionskip
  \sbox\@tempboxa{\small #1. #2}%
  \ifdim \wd\@tempboxa >\hsize
    \small #1. #2\par
  \else
    \global \@minipagefalse
    \hb@xt@\hsize{\hfil\box\@tempboxa\hfil}%
  \fi
  \vskip\belowcaptionskip}

\renewenvironment{thebibliography}[1]{%
\begin{oldthebibliography}{#1}%
\small%
\raggedright%
\setlength{\itemsep}{5pt plus 0.2ex minus 0.05ex}%
}%
{%
\end{oldthebibliography}%
}

\makeatother

\usepackage[T1]{fontenc} 
\usepackage{romannum} 

\begin{document} 


\title{{\boldmath Heat kernel approach for confined quantum gas} \\  }


\author[a,c]{Ping Zhang} 
\author[b,c*]{and Tong Liu} \note{liutong@tju.edu.cn}


\affiliation[a]{School of Finance, Capital University of Economics and Business, Beijing 100070, P. R. China}
\affiliation[b]{School of Architecture, Tianjin University, Tianjin 300072, P. R. China}
\affiliation[c]{Department of Physics, Tianjin University, Tianjin 300350, P.R. China}

\abstract{In this paper, based on the heat kernel technique, we calculate equations of
state and thermodynamic quantities for ideal quantum gases in confined space
with external potential. Concretely, we provide expressions for equations of
state and thermodynamic quantities by means of heat kernel coefficients for
ideal quantum gases. Especially, using an analytic continuation treatment, we
discuss the application of the heat kernel technique to Fermi gases in which
the expansion diverges when the fugacity $z>1$. In order to calculate the modification of heat kernel coefficients caused by external potentials, we suggest
an approach for calculating the expansion of the global heat kernel of the
operator $-\Delta+U\left(  x\right)  $ based on an approximate method of the
calculation of spectrum in quantum mechanics. At last, we discuss the
properties of quantum gases under the condition of weak and complete
degeneration, respectively.}




\maketitle 

\flushbottom

\section{Introduction}

In this paper, we consider ideal quantum gases in confined space with external
potentials. First, we provide the expansion of equations of state for ideal
Bose and Fermi gases by means of heat kernel coefficients. Such a result
allows us to calculate the thermodynamic properties of ideal quantum gases
with the help of the heat kernel technique which has been studied thoroughly
in mathematics and physics. Second, for the purpose of calculating the
thermodynamic quantities in confined space with the heat kernel technique,
based on an approximate method of the calculation of spectra
\cite{kirsch2011introduction,chadan1997introduction,amore2010can}, we suggest
an approach to calculate the modification of the global heat kernel caused by
an external potential. Third, we calculate the influence of boundaries and
external potentials to the behavior of ideal Bose and Fermi gases based on the
heat kernel method.

Especially, in this paper, we will deal with the Fermi case by the heat kernel
method. For Bose cases, some authors have used the heat kernel method to the
calculation of the thermodynamic behaviors of ideal Bose gases
\cite{accikkalp2015application}. For the Fermi case, however, the expansion of
the thermodynamic quantities will diverge when the fugacity becomes greater
than $1$. Using an analytic continuation treatment, we give a heat kernel
expansion to the thermodynamic behaviors of ideal Fermi gases. The result
shows that though the series of an expansion of a Fermi thermodynamic quantity
will diverge when $z>1$, one still can achieve a finite result by analytic continuation.

The basis of this paper is the heat kernel technique. The local heat kernel
$K\left(  t;x,y\right)  $ is the Green function of the initial-value problem
of the heat-type equation $\left(  \frac{\partial}{\partial t}+\mathcal{D}%
\right)  \psi=0$. The global heat kernel $K\left(  t\right)  $ is the trace of
the local heat kernel:
\begin{equation}
K\left(  t\right)  =\int d^{D}xK\left(  t;x,x\right)  =\sum_{n}e^{-\lambda
_{n}t}, \label{Heat Kernel 1}%
\end{equation}
where $x,y$ are coordinates of the $D$-dimensional space and $\lambda_{n}$ is
the eigenvalue of the operator $\mathcal{D}$. For a second-order differential
operator of Laplace-type with a local boundary, the corresponding
$D$-dimensional global heat kernel $K\left(  t\right)  $ can be asymptotically
expanded as \cite{vassilevich2003heat}
\begin{equation}
K\left(  t\right)  =\left(  4\pi t\right)  ^{-D/2}\sum_{l=0,\frac{1}%
{2},1,\cdots}^{\infty}B_{l}t^{l}, \label{Heat Kernel 2}%
\end{equation}
where $B_{l}$ is the heat kernel coefficient. In two-dimensional cases, the
first important result given by Weyl shows that the leading term of the heat
kernel expansion is proportional to the area \cite{weyl1968gesammelte}. Then,
Pleijel proved that the second term of the global heat kernel is proportional
to the perimeter \cite{pleijel1954study}. Moreover, Kac hypothesized that the
third term is proportional to the Euler-Poincar\'{e} characteristic number
\cite{kac1966can}. Recently, there are many researches on the calculation of
heat kernels and the corresponding physical quantities
\cite{dai2009number,dai2010approach}.

The thermodynamic properties of gases in confined space have been widely
investigated in recent years, including classical gases
\cite{sisman2004casimir,sisman2004surface,guo2012performance} and quantum
gases \cite{dai2003quantum,dai2004geometry,pang2006difference}. Many studies
show that in confined space, ideal gases will show non-uniform
\cite{mukherjee2018fisher,aydin2019quantum} or anisotropy
\cite{pang2011pressure} due to the existence of the boundary. In addition, the
studies on the influence of boundary to ideal gases lead to some other
progresses. For weak interaction quantum gases, the interaction between
particles can be treated as some kinds of boundary effects, and the behaviors
of non-ideal quantum gases are predicted
\cite{dai2005hard,dai2007interacting,dai2007upper}. Also based on the studies
on boundary effects, the problems of thermodynamic cycles and heat engines
draw many attentions recently
\cite{karabetoglu2017thermosize,aydin2019quantum,nie2009performance,nie2008micro,nie2009quantum,accikkalp2016assessment}%
.

In this paper, we calculate the influence of boundaries and external
potentials to the thermodynamic properties of ideal Bose and Fermi gases.
First, we express equations of state and thermodynamic quantities by means of
the heat kernel coefficients. With the help of heat kernel expansion, Eq.
(\ref{Heat Kernel 2}), we acquire the expression of the equations of state and
thermodynamic quantities by means of heat kernel coefficients. Next, we
introduce a method for calculating the modification to the heat kernel
coefficient. Because the effect of the boundary and the potential is reflected
in the heat kernel coefficients, we develop a method to calculate the heat
kernel coefficients through the method of the approximate spectrum provided by
Refs. \cite{kirsch2011introduction,chadan1997introduction,amore2010can}. At
last, we analyze the properties of ideal Bose and Fermi gases under weak
degenerate and completely degenerate conditions, respectively. Note that in
principle a finite size system should be considered in canonical ensembles
\cite{zhou2018canonical}. A comparison between canonical ensembles and grand
canonical ensembles is provided in Ref. \cite{pang2006difference}.

In section 2, we achieve the expression of the partition function, the grand
potential and the corresponding thermodynamic quantities by means of the heat
kernel coefficients of ideal Bose and Fermi gases in quantum statistics. In
section 3, we calculate the modification of the global heat kernel by the
spectrum asymptotic method in one-dimensional space. In section 4, we
introduce a method for calculating the modification of the heat kernel
coefficient in $D$-dimensional $\left(  D\geq2\right)  $ confined space. In
section 5, we calculate the effect of boundary and potential. Specifically, we
discuss the properties of ideal Bose and Fermi gases under the conditions of
weak degeneration and complete degeneration, respectively. The conclusions are
drawn in the section 6.

\section{Expansions of partition functions and grand potentials: the heat
kernel method}

In this section, we provide an expansion for partition functions, grand
potentials, and the corresponding thermodynamic quantities by means of heat
kernel coefficients.

\subsection{Expressions of partition functions and grand potentials with heat
kernel coefficients}

Comparing the partition function of a classical ideal gas
\begin{equation}
Z\left(  \beta\right)  =\sum_{s}\exp\left(  -\beta E_{s}\right)
\end{equation}
with the definition of the global heat kernel, Eq. (\ref{Heat Kernel 1}), and
using the heat kernel expansion, Eq. (\ref{Heat Kernel 2}), we can expand the
partition function as
\begin{align}
Z  &  =K\left(  \frac{\lambda^{2}}{4\pi}\right) \nonumber\\
&  =\sum_{l=0,\frac{1}{2},1,\cdots}^{\infty}\frac{B_{l}}{\lambda^{D-2l}%
}\left(  \frac{1}{4\pi}\right)  ^{l}, \label{HKexpansion}%
\end{align}
where $\lambda=\frac{h}{\sqrt{2\pi mkT}}$ is the mean thermal wavelength and
$B_{l}$ is the heat kernel coefficient of the operator $-\Delta+\frac
{2m}{\hbar^{2}}V$.

In quantum statistics, the grand potential of ideal Bose and Fermi gases is
\begin{equation}
\ln\Xi=\mp\sum_{s}\ln\left(  1\mp ze^{-\beta E_{s}}\right)  .
\end{equation}
In this equation and following, the upper sign represents bosons and the lower
sign represents fermions. The grand potential can be expanded as a series of
global heat kernel:
\begin{align}
\ln\Xi &  =\mp\sum_{s}\sum_{n=1}^{\infty}\frac{\left(  \pm1\right)  ^{n+1}}%
{n}e^{-n\beta E_{s}}z^{n}\nonumber\\
&  =\sum_{n=1}^{\infty}\frac{\left(  \pm1\right)  ^{n+1}}{n}K\left(
n\frac{\lambda^{2}}{4\pi}\right)  z^{n}. \label{sum1}%
\end{align}
Substituting Eq. (\ref{HKexpansion}) into Eq. (\ref{sum1}) gives%
\begin{equation}
\ln\Xi=\sum_{l=0,\frac{1}{2},1,\cdots}^{\infty}\left(  \frac{1}{4\pi}\right)
^{l}\lambda^{2l-D}B_{l}\sum_{n=1}^{\infty}\left(  \pm1\right)  ^{n+1}%
n^{l-D/2-1}z^{n}. \label{sum}%
\end{equation}

The summation over $n$ in Eq. (\ref{sum}) for Bose and Fermi case are
significant difference: the sum converges for the Bose case, but diverges for
the Fermi case when $z>1$.

For the Fermi case, the convergence of the summation $\sum_{n=1}^{\infty
}\left(  -1\right)  ^{n+1}n^{l-D/2-1}z^{n}$ requires $z<1$. However, the range
of value of $z$ in the Fermi case is $0<z<\infty$. To understand why the
summation diverges for $z>1$, we first perform the summation in the case of
$z<1$. For $z<1$, performing the summation gives $\sum_{n=1}^{\infty}\left(
\pm1\right)  ^{n+1}n^{l-D/2-1}z^{n}=f_{D/2+1-l}\left(  z\right)  $, where
$f_{\nu}\left(  z\right)  $ is the Fermi-Dirac integral. The reason why the
radius of convergence of this series is $1$ can be understand by analyzing the
analyticity properties of the function $f_{\nu}\left(  z\right)  $ in the $z-$plane.

\begin{figure}[ptb]
\centering
\includegraphics[width=4in]{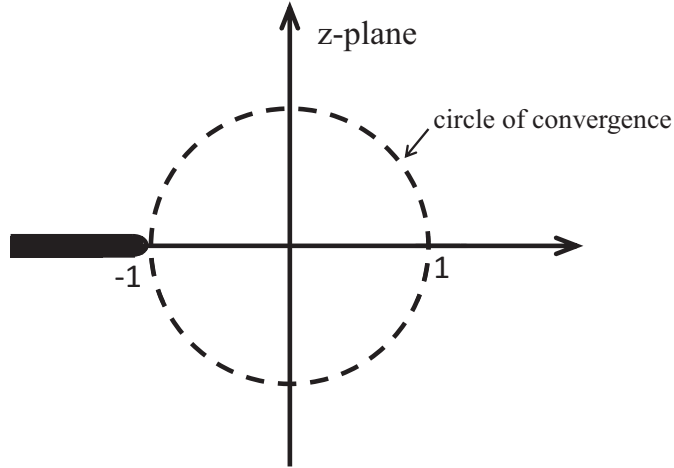}\caption{The analyticity area of the
Fermi-Dirac integral $f_{\nu}\left(  z\right)  $ and the circle of convergence
of the series in Eq. (\ref{sum}).}%
\end{figure}

The analyticity area of the Fermi-Dirac integral $f_{\nu}\left(  z\right)  $
is shown in Fig. 1 \cite{leonard1968exact,dai2009exactly,dai2017explicit}.
From Fig. 1 we can see that the singularity in the $z-$plane of $f_{\nu
}\left(  z\right)  $ begins with $-1$ to $-\infty$. Therefore, the radius of
the circle of convergence of the series is $1$. However, we can also see from
Fig. 1 that all points on the positive horizontal axis are analytical points.
This implies that we can analytically continue $\sum_{n=1}^{\infty}\left(
\pm1\right)  ^{n+1}n^{l-D/2-1}z^{n}$ to whole positive horizontal axis; in
other word, we can achieve an finite result of the sum of the series.
Consequently, the result for the Fermi case is finite in the region
$-1<z<\infty$. Then we have%
\begin{equation}
\ln\Xi_{Fermi}=\sum_{l}B_{l}\frac{1}{\left(  4\pi\right)  ^{l}}\lambda
^{2l-D}f_{D/2+1-l}\left(  z\right)  .
\end{equation}
This result is finite for any non-negative $z$.

For the Bose case, the sum in Eq. (\ref{sum}) is converge and the summation
can be done directly; the result for the Bose case has been obtained in Refs.
\cite{kirsten1999bose}:%
\begin{equation}
\ln\Xi_{Bose}=\sum_{l}B_{l}\frac{1}{\left(  4\pi\right)  ^{l}}\lambda
^{2l-D}g_{D/2+1-l}\left(  z\right)  ,
\end{equation}
where $g_{\nu}\left(  z\right)  $ is the Bose-Einstein integral.

For convenience, in the following we express the grand potential of both Bose
and Fermi gases as%
\begin{equation}
\ln\Xi=\sum_{l=0,\frac{1}{2},1,\cdots}^{\infty}B_{l}\lambda^{2l-D}\frac
{1}{\left(  4\pi\right)  ^{l}}h_{D/2+1-l}\left(  z\right)  , \label{lnXi}%
\end{equation}
where $h_{\nu}\left(  z\right)  =\frac{1}{\Gamma\left(  \nu\right)  }\int%
_{0}^{\infty}\frac{t^{\nu-1}}{z^{-1}e^{t}\mp1}dt$ equals to Bose-Einstein
integral $g_{\nu}\left(  z\right)  $ and Fermi-Dirac integral $f_{\nu}\left(
z\right)  $, respectively.

\subsection{The expression of the equation of state and thermodynamic
quantities}

In this section, we give the expression of the equations of state and the
corresponding thermodynamic quantities in quantum statistics.

According to Eq. (\ref{lnXi}), the equation of state is%
\begin{equation}
\left\{
\begin{array}
[c]{c}%
\displaystyle\frac{pV}{kT}=\ln\Xi=\sum_{l=0,\frac{1}{2},1,\cdots}^{\infty
}\frac{B_{l}}{\lambda^{D-2l}}\left(  \frac{1}{4\pi}\right)  ^{l}%
h_{D/2+1-l}\left(  z\right)  ,\\
\displaystyle N=z\frac{\partial}{\partial z}\ln\Xi=\sum_{l=0,\frac{1}%
{2},1,\cdots}^{\infty}\frac{B_{l}}{\lambda^{D-2l}}\left(  \frac{1}{4\pi
}\right)  ^{l}h_{D/2-l}\left(  z\right)  .
\end{array}
\right.  \label{EOS fermi/bose}%
\end{equation}
From Eq. (\ref{EOS fermi/bose}), we can directly obtain various thermodynamic
quantities: the internal energy%
\begin{equation}
\frac{U}{kT}=-\beta\frac{\partial}{\partial\beta}\ln\Xi=\sum_{l=0,\frac{1}%
{2},1,\cdots}^{\infty}\frac{B_{l}}{\lambda^{D-2l}}\left(  \frac{1}{4\pi
}\right)  ^{l}\left(  \frac{D}{2}-l\right)  h_{D/2+1-l}\left(  z\right)  ,
\label{U fermi/bose}%
\end{equation}
the Helmholtz free energy%
\begin{align}
\frac{F}{kT}  &  =-\ln\Xi+\frac{\mu}{kT}N\nonumber\\
&  =\sum_{l=0,\frac{1}{2},1,\cdots}^{\infty}\frac{B_{l}}{\lambda^{D-2l}%
}\left(  \frac{1}{4\pi}\right)  ^{l}\left[  h_{D/2-l}\left(  z\right)  \ln
z-h_{D/2+1-l}\left(  z\right)  \right]  ,
\end{align}
the entropy%
\begin{align}
\frac{S}{k}  &  =\frac{U-F}{kT}\nonumber\\
&  =-N\ln z+\sum_{l=0,\frac{1}{2},1,\cdots}^{\infty}\frac{B_{l}}%
{\lambda^{D-2l}}\left(  \frac{1}{4\pi}\right)  ^{l}\left[  \left(  \frac{D}%
{2}+1-l\right)  h_{D/2+1-l}\left(  z\right)  \right]  ,
\end{align}
and the specific heat
\begin{align}
&  \frac{C_{V}}{k}=\frac{1}{k}\left(  \frac{\partial U}{\partial T}\right)
_{V}\nonumber\\
&  =\sum_{l=0,\frac{1}{2},1,\cdots}^{\infty}\frac{B_{l}}{\lambda^{D-2l}%
}\left[  \frac{D^{2}}{4}+\frac{D}{2}-\left(  D+1\right)  l+l^{2}\right]
\left(  \frac{1}{4\pi}\right)  ^{l}h_{D/2+1-l}\left(  z\right) \nonumber\\
&  \displaystyle-\frac{\left\{  \sum_{l=0,\frac{1}{2},1,\cdots}^{\infty
}\left(  B_{l}/\lambda^{D-2l}\right)  \left(  \frac{D}{2}-l\right)  \left[
1/\left(  4\pi\right)  \right]  ^{l}h_{D/2-l}\left(  z\right)  \right\}  ^{2}%
}{\sum_{s=0,\frac{1}{2},1,\cdots}^{\infty}\left(  B_{l}/\lambda^{D-2s}\right)
\left[  1/\left(  4\pi\right)  \right]  ^{s}h_{D/2-1-s}\left(  z\right)  },
\label{Cv fermi/bose}%
\end{align}
where the relation
\begin{equation}
\frac{\partial z}{\partial T}=\displaystyle-\frac{z}{T}\frac{\sum
_{l=0,\frac{1}{2},1,\cdots}^{\infty}\left(  \frac{D}{2}-l\right)  B_{l}\left[
\lambda^{2}/\left(  4\pi\right)  \right]  ^{l}h_{D/2-l}\left(  z\right)
}{\sum_{s=0,\frac{1}{2},1,\cdots}^{\infty}B_{s}\left[  \lambda^{2}/\left(
4\pi\right)  \right]  ^{s}h_{D/2-1-s}\left(  z\right)  }%
\end{equation}
is used when calculating the specific heat $C_{V}$.

\section{Spectra and heat kernel coefficients in one-dimensional confined
space with potentials}

In this section, we calculate an approximate spectrum of a particle in
one-dimensional confined space with an external potential. Moreover, we give
the corresponding modified heat kernel coefficients. Some external potentials
are considered.

\subsection{Spectra: one dimension}

The one-dimensional Schr\"{o}dinger equation with the Dirichlet boundary
condition is
\begin{equation}
\left\{
\begin{array}
[c]{l}%
\displaystyle\left(  -\frac{\hbar^{2}}{2m}\frac{d^{2}}{dx^{2}}+V\left(
x\right)  \right)  \psi\left(  x\right)  =E\psi\left(  x\right)  ,\\
\displaystyle\psi\left(  0\right)  =\psi\left(  L\right)  =0,
\end{array}
\right.
\end{equation}
where $0<x<L$. According to Refs.
\cite{kirsch2011introduction,chadan1997introduction}, we express the
approximate spectrum as
\begin{equation}
E_{n}\simeq\frac{\pi^{2}\hbar^{2}}{2mL^{2}}\left[  n^{2}+\int_{0}^{1}U\left(
\xi\right)  d\xi-\int_{0}^{1}U\left(  \xi\right)  \cos\left(  2\pi
n\xi\right)  d\xi\right]  ,\text{ }n=1,2,3,\cdots, \label{En V(xi)}%
\end{equation}
where $U\left(  \xi\right)  =\frac{2mL^{2}}{\pi^{2}\hbar^{2}}V\left(
L\xi\right)  $.

To give a more explicit expression to Eq. (\ref{En V(xi)}), we write the
potential $U\left(  \xi\right)  $ as
\begin{equation}
U\left(  \xi\right)  =A\left(  \xi\right)  +S\left(  \xi\right)  ,
\end{equation}
where the function $A\left(  \xi\right)  $ is an odd function with respect to
the axis $\xi=\frac{1}{2}$ and $S\left(  \xi\right)  $ is an even one. The
relations between $A\left(  \xi\right)  $, $S\left(  \xi\right)  $ and
$U\left(  \xi\right)  $ are%
\begin{equation}
\left\{
\begin{array}
[c]{c}%
A\left(  \xi\right)  =\frac{1}{2}\left[  U\left(  \xi\right)  -U\left(
1-\xi\right)  \right] \\
S\left(  \xi\right)  =\frac{1}{2}\left[  U\left(  \xi\right)  +U\left(
1-\xi\right)  \right]
\end{array}
\right.  . \label{A S}%
\end{equation}
Because of $\int_{0}^{1}A\left(  \xi\right)  d\xi=\int_{0}^{1}A\left(
\xi\right)  \cos\left(  2\pi n\xi\right)  d\xi=0$, we have%
\begin{equation}
\int_{0}^{1}U\left(  \xi\right)  d\xi=\int_{0}^{1}S\left(  \xi\right)  d\xi,
\label{U}%
\end{equation}
and%
\begin{equation}
\int_{0}^{1}U\left(  \xi\right)  \cos\left(  2\pi n\xi\right)  d\xi=\int%
_{0}^{1}S\left(  \xi\right)  \cos\left(  2\pi n\xi\right)  d\xi. \label{Ucos}%
\end{equation}
That is to say, only the even part $S\left(  \xi\right)  $ in the potential
$U\left(  \xi\right)  $ contributes to the energy spectrum $E_{n}$ under the
approximation, Eq. (\ref{En V(xi)}).

The even part $S\left(  \xi\right)  $ can always be expanded as
\begin{equation}
S\left(  \xi\right)  =\sum_{m=0}^{\infty}a_{m}\left(  \xi-\frac{1}{2}\right)
^{2m}. \label{S(xi)}%
\end{equation}
Substituting Eq. (\ref{S(xi)}) into Eqs. (\ref{U}) and (\ref{Ucos}) and
integrating term by term give
\begin{equation}
\int_{0}^{1}U\left(  \xi\right)  d\xi=\sum_{m=0}^{\infty}a_{m}\int_{0}%
^{1}\left(  \xi-\frac{1}{2}\right)  ^{2m}d\xi=\sum_{m=0}^{\infty}\frac{a_{m}%
}{2^{2m}\left(  2m+1\right)  }, \label{ingral V}%
\end{equation}
and%
\begin{align}
&  \int_{0}^{1}U\left(  \xi\right)  \cos\left(  2\pi n\xi\right)
d\xi\nonumber\\
&  =\sum_{m=0}^{\infty}a_{m}\int_{0}^{1}\left(  \xi-\frac{1}{2}\right)
^{2m}\cos\left(  2\pi n\xi\right)  d\xi\nonumber\\
&  =\sum_{m=0}^{\infty}a_{m}\left(  -1\right)  ^{n+1}4^{-m}\frac{1}{2}\left[
\mathcal{E}_{-2m}\left(  -i\pi n\right)  +\mathcal{E}_{-2m}\left(  i\pi
n\right)  \right]  , \label{integral V cos}%
\end{align}
where $\mathcal{E}_{n}\left(  z\right)  =\int_{1}^{\infty}e^{-zt}t^{-n}dt$ is
the exponential integral function. When $n\rightarrow\infty$,
\begin{equation}
\left(  -1\right)  ^{n+1}4^{-m}\frac{1}{2}\left[  \mathcal{E}_{-2m}\left(
-i\pi n\right)  +\mathcal{E}_{-2m}\left(  i\pi n\right)  \right]
\simeq2^{1-2m}\frac{m}{n^{2}\pi^{2}},
\end{equation}
we then have%
\begin{equation}
\int_{0}^{1}U\left(  \xi\right)  \cos\left(  2\pi n\xi\right)  d\xi\simeq
\sum_{m=0}^{\infty}a_{m}\frac{m}{2^{2m-1}\pi^{2}n^{2}}.
\label{integral V cos 1}%
\end{equation}

Substituting Eqs. (\ref{ingral V}) and (\ref{integral V cos 1}) into Eq.
(\ref{En V(xi)}), we obtain a more explicit expression of energy spectrum to
an arbitrary potential:%
\begin{equation}
E_{n}\simeq\frac{\pi^{2}\hbar^{2}}{2mL^{2}}\left(  n^{2}+\frac{\tilde{W}%
}{n^{2}}+\tilde{U}\right)  ,\text{ \ \ }n=1,2,3,\cdots, \label{En V(a,b)}%
\end{equation}
where
\begin{equation}
\tilde{U}=\sum_{m=0}^{\infty}\frac{a_{m}}{2^{2m}\left(  2m+1\right)  }
\label{a}%
\end{equation}
and%
\begin{equation}
\tilde{W}=-\frac{1}{\pi^{2}}\sum_{m=1}^{\infty}a_{m}\frac{m}{2^{2m-1}}.
\label{b}%
\end{equation}

\subsection{Heat kernel coefficients: one dimension}

With the energy spectrum Eq. (\ref{En V(a,b)}), we can directly express the
global heat kernel as
\begin{equation}
K\left(  t\right)  \simeq\exp\left(  -t\frac{\pi^{2}}{L^{2}}\tilde{U}\right)
\sum_{n=1}^{\infty}\exp\left[  -t\frac{\pi^{2}}{L^{2}}\left(  n^{2}%
+\frac{\tilde{W}}{n^{2}}\right)  \right]  .
\end{equation}
Using the Euler-Maclaurin formula \cite{arfken2012mathematical}
\begin{align}
\sum_{n=a}^{b}f\left(  n\right)   &  =\int_{a}^{b}f\left(  n\right)
dn+\frac{1}{2}\left[  f\left(  a\right)  +f\left(  b\right)  \right]
\nonumber\\
&  +\sum_{k=1}^{m}\frac{\mathcal{B}_{2k}}{\left(  2k\right)  !}\left[
f^{\left(  2k-1\right)  }\left(  b\right)  -f^{\left(  2k-1\right)  }\left(
a\right)  \right] \nonumber\\
&  +\frac{1}{\left(  2m+1\right)  !}\int_{a}^{b}\mathcal{B}_{2m+1}\left(
x-\left\lfloor x\right\rfloor \right)  f^{\left(  2m+1\right)  }\left(
x\right)  dx,
\end{align}
where $\mathcal{B}_{2k}$ is Bernoulli number, $\mathcal{B}_{2k}\left(
x\right)  $ is Bernoulli polynomial, and $\left\lfloor x\right\rfloor $ is the
maximum integer which is no more than $x$, taking $m=1$ and omitting the
remainder term, we obtain%
\begin{align}
K\left(  t\right)   &  \simeq\frac{1}{\sqrt{4\pi t}}\exp\left(  -t\frac
{\pi^{2}}{L^{2}}\tilde{U}\right)  \times\nonumber\\
&  \left\{  \frac{L}{2}\left\{  \exp\left(  -2\frac{\pi^{2}}{L^{2}}%
\sqrt{\tilde{W}}t\right)  \operatorname{erfc}\left[  \frac{\pi}{L}\left(
1-\sqrt{\tilde{W}}\right)  \sqrt{t}\right]  \right.  \right. \nonumber\\
&  \left.  +\exp\left(  2\frac{\pi^{2}}{L^{2}}\sqrt{\tilde{W}}t\right)
\operatorname{erfc}\left[  \frac{\pi}{L}\left(  1+\sqrt{\tilde{W}}\right)
\sqrt{t}\right]  \right\} \nonumber\\
&  \left.  +\frac{\sqrt{4\pi t}}{6}\left[  3+t\frac{\pi^{2}}{L^{2}}\left(
1-\tilde{W}\right)  \right]  \exp\left[  -t\frac{\pi^{2}}{L^{2}}\left(
1+\tilde{W}\right)  \right]  \right\}  , \label{K(t) D=1}%
\end{align}
where $\operatorname{erfc}\left(  z\right)  $ is the complementary error
function \cite{brychkov2008handbook}.

Expanding the global heat kernel Eq. (\ref{K(t) D=1}) as the form of Eq.
(\ref{Heat Kernel 2}) gives the heat kernel coefficients:
\begin{equation}
B_{0}=L, \label{B0 1D}%
\end{equation}%
\begin{equation}
B_{1/2}=\pi^{1/2}\left(  1-2\tilde{W}^{1/2}\right)  , \label{B1/2 1D}%
\end{equation}%
\begin{equation}
B_{1}=-\frac{\pi^{2}}{L}\tilde{U}, \label{B1 1D}%
\end{equation}%
\begin{equation}
B_{3/2}=-\frac{\pi^{5/2}}{3L^{2}}\left[  2+3\tilde{U}+6\left(  1-\tilde
{U}\right)  \tilde{W}^{1/2}+4\tilde{W}-2\tilde{W}^{3/2}\right]  ,
\label{B3/2 1D}%
\end{equation}%
\begin{equation}
B_{2}=\frac{\pi^{4}}{2L^{3}}\left(  \tilde{U}^{2}+4\tilde{W}\right)  ,
\end{equation}%
\begin{align}
B_{5/2}  &  =\frac{\pi^{9/2}}{30L^{4}}\left[  5+20\tilde{U}+15\tilde{U}%
^{2}+10\left(  1+6\tilde{U}-30\tilde{U}^{2}\right)  \tilde{W}^{1/2}\right.
\nonumber\\
&  \left.  +10\left(  3+4\tilde{U}\right)  \tilde{W}-20\left(  3+\tilde
{U}\right)  \tilde{W}^{3/2}+25\tilde{W}^{2}-6\tilde{W}^{5/2}\right]  ,
\end{align}%
\begin{equation}
B_{3}=-\frac{\pi^{6}}{6L^{5}}\left(  \tilde{U}^{3}+12\tilde{U}\tilde
{W}\right)  ,
\end{equation}%
\begin{align}
B_{7/2}  &  =-\frac{\pi^{13/2}}{210L^{6}}\left[  35\left(  \tilde{U}%
+2\tilde{U}^{2}+\tilde{U}^{3}\right)  \right. \nonumber\\
&  \left.  +14\left(  1+5\tilde{U}+15\tilde{U}^{2}-5\tilde{U}^{3}\right)
\tilde{W}^{1/2}+70\left(  1+3\tilde{U}+2\tilde{U}^{2}\right)  \tilde{W}\right.
\nonumber\\
&  \left.  +70\left(  3-6\tilde{U}-\tilde{U}^{2}\right)  \tilde{W}%
^{3/2}+35\left(  4+5\tilde{U}\right)  \tilde{W}^{2}\right. \nonumber\\
&  \left.  -14\left(  5+3\tilde{U}\right)  \tilde{W}^{5/2}+70\tilde{W}%
^{3}-10\tilde{W}^{7/2}\right]  ,
\end{align}%
\begin{equation}
B_{4}=\frac{\pi^{8}}{24L^{7}}(\tilde{U}^{4}+24\tilde{U}^{2}\tilde{W}%
+16\tilde{W}^{2}),
\end{equation}%
\begin{align}
B_{9/2}  &  =\frac{\pi^{17/2}}{7560L^{8}}\left[  105\left(  -1+6\tilde{U}%
^{2}+8\tilde{U}^{3}+3\tilde{U}^{4}\right)  \right. \nonumber\\
&  \left.  18\left(  5+28\tilde{U}+70\tilde{U}^{2}+140\tilde{U}^{3}%
-35\tilde{U}^{4}\right)  \tilde{W}^{1/2}\right. \nonumber\\
&  \left.  420\left(  1+6\tilde{U}+9\tilde{U}^{2}+4\tilde{U}^{3}\right)
\tilde{W}\right. \nonumber\\
&  \left.  +840\left(  1+9\tilde{U}-9\tilde{U}^{2}-\tilde{U}^{3}\right)
\tilde{W}^{3/2}\right. \nonumber\\
&  \left.  +630\left(  3+8\tilde{U}+5\tilde{U}^{2}\right)  \tilde{W}%
^{2}-252\left(  15+10\tilde{U}+3\tilde{U}^{2}\right)  \tilde{W}^{5/2}\right.
\nonumber\\
&  \left.  +420(5+6\tilde{U})\tilde{W}^{3}-72(7+5\tilde{U})\tilde{W}%
^{7/2}+735\tilde{W}^{4}-70\tilde{W}^{9/2}\right]  ,
\end{align}%
\begin{equation}
B_{5}=-\frac{\pi^{10}}{120L^{9}}\left(  \tilde{U}^{5}+40\tilde{U}^{3}\tilde
{W}+80\tilde{U}\tilde{W}^{2}\right)  ,
\end{equation}
and so on. We find that the modification of the heat kernel coefficients
caused by the term $\int_{0}^{1}U\left(  \xi\right)  d\xi$ and $\int_{0}%
^{1}U\left(  \xi\right)  \cos\left(  2\pi n\xi\right)  d\xi$ begins at $B_{1}$
and $B_{3/2}$, respectively.

\subsection{Various external potentials: examples}

In the following, as examples, we consider various external potentials in
confined space.

(1) For the external potential
\[
U\left(  \xi\right)  =\frac{U_{0}}{\sqrt{2\pi}\sigma}\exp\left[  -\frac
{1}{2\sigma^{2}}\left(  \xi-\frac{1}{2}\right)  ^{2}\right]  ,
\]
the coefficient in Eqs. (\ref{a}) and (\ref{b}) reads%
\begin{equation}
a_{m}=U_{0}\frac{2^{-1/2-m}\sigma^{-1-2m}}{\sqrt{\pi}m!}.
\end{equation}
By Eqs. (\ref{a}), (\ref{b}), and (\ref{S(xi)}), we achieve%
\begin{equation}
\tilde{U}=U_{0}\operatorname{erfi}\left(  \frac{1}{2\sqrt{2}\sigma}\right)
\text{ and\ }\tilde{W}=-U_{0}\frac{e^{1/\left(  8\sigma^{2}\right)  }}%
{4\sqrt{2}\pi^{5/2}\sigma^{3}},
\end{equation}
where $\operatorname{erfi}\left(  z\right)  $ is the imaginary error function
\cite{brychkov2008handbook}.

(2) For the external potential
\begin{equation}
U\left(  \xi\right)  =U_{0}\left\{  \left[  \left(  \xi-\frac{1}{2}\right)
^{2}-\frac{1}{8}\alpha^{2}\right]  ^{2}-\frac{1}{64}\alpha^{4}\right\}
,\text{ \ \ }\left(  0<\alpha<1\right)  ,
\end{equation}
we have%
\begin{equation}
a_{0}=0,\text{ \ \ }a_{1}=-\frac{\alpha^{2}}{4}U_{0},\text{ \ \ }a_{2}%
=U_{0},\text{ \ \ }a_{m}=0\text{ \ \ }\left(  m\geq3\right)  ,
\end{equation}
and then achieve%
\begin{equation}
\tilde{U}=U_{0}\left(  \frac{1}{80}-\frac{\alpha^{2}}{48}\right)  \text{
and\ }\tilde{W}=\frac{U_{0}}{8\pi^{2}}\left(  \alpha^{2}-2\right)  .
\end{equation}

(3) For the external potential
\begin{equation}
U\left(  \xi\right)  =U_{0}\alpha^{2}e^{\alpha^{2}(\xi-1/2)^{2}}\left(
\xi-\frac{1}{2}\right)  ^{2}\left[  1+\alpha^{2}\left(  \xi-\frac{1}%
{2}\right)  ^{2}\right]  ,
\end{equation}
we have
\begin{equation}
a_{m}=\frac{m^{2}}{m!}\alpha^{2m},\text{ \ \ }m=0,1,2,\cdots,
\end{equation}
and then achieve%
\begin{equation}
\tilde{U}=\frac{1}{8}\left[  2\sqrt{\pi}\operatorname{erfi}\left(  \frac{1}%
{2}\right)  -e^{1/4}\right]  \text{ and\ }\tilde{W}=-\frac{29e^{1/4}}%
{32\pi^{2}}.
\end{equation}

\section{Spectra and heat kernel coefficients in $D\left(  \geq2\right)
$-dimensional confined space}

In this section, the ideal gas in the $D$-dimensional $\left(  D\geq2\right)
$ confined space is discussed. We calculate the modification of external
potentials to global heat kernel in confined space. The method given by Ref.
\cite{amore2010can} is used to acquire the approximate energy spectrum.
Specifically, we provide the modified heat kernel coefficients in
two-dimensional confined space and three-dimensional confined ball. Some
external potentials are considered.

In fact, the heat kernel technique is a high frequency asymptotics method. The
method provided by Ref. \cite{amore2010can} achieves the asymptotics of
eigenvalues in confined space. The key step in Ref. \cite{amore2010can} is to
take a high frequency approximation to the matrix element. Such an
approximation reveals what essentially one has done in Weyl and Kac's famous
high frequency asymptotics for the heat kernel, although it is just equivalent
to substituting an external potential with its meanvalue in the confined space.

\subsection{Spectra and heat kernel coefficients: $D\left(  \geq2\right)  $
dimensions}

According to Ref. \cite{amore2010can}, the approximate eigenvalues of the
equation%
\begin{equation}
\left\{
\begin{array}
[c]{l}%
\displaystyle\left(  -\frac{\hbar^{2}}{2m}\Delta+U\right)  \psi=E\psi,\text{
\ \ in \ \ }\Omega\\
\displaystyle\psi=0,\text{ \ \ on \ \ }\partial\Omega
\end{array}
\right.
\end{equation}
are
\begin{equation}
E_{s}\simeq E_{s}^{\left(  0\right)  }+\tilde{U}, \label{spectral}%
\end{equation}
where
\begin{equation}
\tilde{U}=\frac{1}{V}\int_{\Omega}U\left(  x\right)  dV
\end{equation}
and $\Omega$ represents the region of space.

Obviously, the global heat kernel in $D$-dimensional $\left(  D\geq2\right)  $
confined space is%
\begin{align}
K\left(  t\right)   &  =\sum_{s}\exp\left(  -t\frac{2mE_{s}}{\hbar^{2}}\right)
\nonumber\\
&  =\left(  4\pi t\right)  ^{-D/2}\sum_{l=0,\frac{1}{2},1,\cdots}^{\infty
}B_{l}t^{l}, \label{K(t)}%
\end{align}
where $B_{l}$ is the heat kernel coefficient of the operator $-\Delta
+\frac{2m}{\hbar^{2}}V$. Substituting Eq. (\ref{spectral}) into Eq.
(\ref{K(t)}), we have%
\begin{align}
K\left(  t\right)   &  \simeq\sum_{s}\exp\left[  -t\frac{2m}{\hbar^{2}}\left(
E_{s}^{\left(  0\right)  }+\tilde{U}\right)  \right] \nonumber\\
&  =\exp\left(  -t\frac{2m\tilde{U}}{\hbar^{2}}\right)  \sum_{s}\exp\left(
-\frac{2mE_{s}^{\left(  0\right)  }}{\hbar^{2}}t\right) \nonumber\\
&  =\exp\left(  -t\tilde{U}\right)  K^{\left(  0\right)  }\left(  t\right)  ,
\label{K(t) 1}%
\end{align}
where $K^{\left(  0\right)  }\left(  t\right)  =\left(  4\pi t\right)
^{-D/2}\sum_{\kappa=0,\frac{1}{2},1,\cdots}^{\infty}B_{\kappa}^{\left(
0\right)  }t^{\kappa}$ is the global heat kernel of $-\Delta$ in the region
$\Omega$. Substituting $\exp\left(  -t\frac{2m\tilde{U}}{\hbar^{2}}\right)
=\sum_{\sigma=0}^{\infty}\frac{\left(  -1\right)  ^{\sigma}}{\sigma!}\left(
\frac{2m\tilde{U}}{\hbar^{2}}\right)  ^{\sigma}t^{\sigma}$ into Eq.
(\ref{K(t) 1}), we obtain the modified heat kernel coefficient%
\begin{equation}
B_{l}=\sum_{s=0}^{\left\lfloor l\right\rfloor }\frac{\left(  -1\right)  ^{s}%
}{s!}\left(  \frac{2m\tilde{U}}{\hbar^{2}}\right)  ^{s}B_{l-s}^{\left(
0\right)  },\text{\ }l=0,\frac{1}{2},1,\cdots, \label{Heat coeff}%
\end{equation}
where $\left\lfloor l\right\rfloor $ is the maximum integer no more than $l$.
The first three coefficients are%
\begin{equation}
B_{0}=B_{0}^{\left(  0\right)  },\text{ \ \ }B_{1/2}=B_{1/2}^{\left(
0\right)  },\text{ \ \ }B_{1}=B_{1}^{\left(  0\right)  }-\frac{2m\tilde{U}%
}{\hbar^{2}}B_{0}^{\left(  0\right)  }. \label{B0 B1/2 B1}%
\end{equation}

\subsection{Heat kernel coefficients in two-dimensional confined space}

The global heat kernel of $-\Delta$ in two-dimensional confined space without
holes, indicated by Kac \cite{kac1966can}, has the asymptotic expression
\begin{equation}
K^{\left(  0\right)  }\left(  t\right)  \simeq\frac{A}{4\pi t}-\frac{L}%
{2\sqrt{\pi t}}+\frac{1}{6},
\end{equation}
where $A$ is the area and $L$ is the perimeter of the region.

When the external potential exists, according to Eq. (\ref{Heat coeff}), the
modified heat kernel coefficients are%
\begin{equation}
B_{0}=A,\text{ }B_{1/2}=-2\sqrt{\pi}L,\text{ }B_{1}=\frac{2\pi}{3}%
-\frac{2m\tilde{U}}{\hbar^{2}}A,
\end{equation}
and the asymptotic expression of the corresponding modified global heat kernel
in two-dimensional confined space is
\begin{equation}
K\left(  t\right)  \simeq\frac{A}{4\pi t}-\frac{L}{2\sqrt{\pi t}}+\left(
\frac{1}{6}-\frac{A}{4\pi}\frac{2m\tilde{U}}{\hbar^{2}}\right)  .
\end{equation}

\subsection{Heat kernel coefficients in a three-dimensional ball}

The global heat kernel of $-\Delta$ in a three-dimensional ball has the
asymptotic expression \cite{bordag1996heat}
\begin{equation}
K^{\left(  0\right)  }\left(  t\right)  \simeq\frac{1}{\left(  4\pi t\right)
^{3/2}}\left(  \frac{4}{3}\pi R^{3}-2\pi^{3/2}R^{2}\sqrt{t}+\frac{8}{3}\pi
Rt\right)  ,
\end{equation}
where $R$ is the radius of the ball.

When there is an external potential in the ball, using Eq. (\ref{Heat coeff}),
we obtain the modified heat kernel coefficients%
\begin{equation}
B_{0}=\frac{4}{3}\pi R^{3},\text{ \ \ }B_{1/2}=-2\pi^{3/2}R^{2},\text{
\ \ }B_{1}=\frac{8}{3}\pi R\left(  1-\frac{R^{2}}{2}\frac{2m\tilde{U}}%
{\hbar^{2}}\right)  . \label{Heat coeff sphere 3D}%
\end{equation}
The corresponding asymptotic expression of the modified global heat kernel in
the three-dimensional ball is
\begin{equation}
K\left(  t\right)  \simeq\frac{1}{\left(  4\pi t\right)  ^{3/2}}\left[
\frac{4}{3}\pi R^{3}-2\pi^{3/2}R^{2}\sqrt{t}+\frac{8}{3}\pi R\left(
1-\frac{R^{2}}{2}\frac{2m\tilde{U}}{\hbar^{2}}\right)  t\right]  .
\end{equation}

\subsection{Various spherically symmetric external potentials in $D\left(
\geq2\right)  $-dimensional balls: examples}

In this section, as examples, we consider some spherically symmetric external
potentials $U\left(  x\right)  $. According to the above analysis, the
modification of the heat kernel coefficients caused by the external potential
$U\left(  x\right)  $ is approximately determined by the quantity $\tilde{U}$.

The effect of a spherically symmetric potential to the energy level in a
$D$-dimensional $\left(  D\geq2\right)  $ ball is%

\begin{align}
\tilde{U}  &  =\frac{1}{V_{ball}}\int_{\Omega}U\left(  r\right)  dV\nonumber\\
&  =\frac{1}{V_{ball}}\int_{0}^{R}U\left(  r\right)  r^{D-1}dr\left(  \int%
_{0}^{2\pi}d\varphi\right)  \left(  \prod\limits_{n=1}^{D-2}\int_{0}^{\pi}%
\sin^{n}\theta d\theta\right) \nonumber\\
&  =\frac{2}{R^{D}}\frac{\Gamma\left(  1+\frac{D}{2}\right)  }{\Gamma\left(
\frac{D}{2}\right)  }\int_{0}^{R}U\left(  r\right)  r^{D-1}dr, \label{U tilde}%
\end{align}
where $R$ is the radius\ of balls, then we have%
\begin{equation}
\tilde{U}=\frac{D}{R^{D}}\int_{0}^{R}U\left(  r\right)  r^{D-1}dr.
\end{equation}

(1) For the external potential%
\begin{equation}
U\left(  r\right)  =-\frac{U_{0}R^{2}}{\sqrt{2\pi}a^{2}}\exp\left(
-\frac{r^{2}}{2a^{2}}\right)  , \label{Gauss}%
\end{equation}
from Eq. (\ref{U tilde}), we achieve
\begin{equation}
\tilde{U}=-U_{0}\frac{D}{\sqrt{\pi}}2^{\left(  D-3\right)  /2}\left(  \frac
{a}{R}\right)  ^{D-2}\left[  \Gamma\left(  \frac{D}{2}\right)  -\Gamma\left(
\frac{D}{2},\frac{R^{2}}{2a^{2}}\right)  \right]  . \label{Gauss Utilde}%
\end{equation}

(2) For the external potential%
\begin{equation}
U=-U_{0}\frac{R^{s}}{r^{s}},\text{ \ \ }s\leq D-1,
\end{equation}
we achieve
\begin{equation}
\tilde{U}=-U_{0}\frac{D}{D-s}.
\end{equation}

(3) For the external potential%
\begin{equation}
U\left(  r\right)  =-U_{0}\frac{\alpha\left(  \alpha-1\right)  }{\cosh
^{2}\left(  r/a\right)  },
\end{equation}
we achieve%
\begin{equation}
\tilde{U}=-\alpha\left(  \alpha-1\right)  U_{0}\frac{R^{2}}{a^{2}}\frac
{D}{R^{D}}\int_{0}^{R}r^{D-1}\operatorname{sech}^{2}\left(  \frac{r}%
{a}\right)  dr.
\end{equation}
When $D=2$,%
\begin{equation}
\tilde{U}=-2\alpha\left(  \alpha-1\right)  U_{0}\left[  \ln\left(
\operatorname{sech}\frac{R}{a}\right)  +\frac{R}{a}\tanh\frac{R}{a}\right]  ;
\end{equation}
when $D=3$,%
\begin{align}
\tilde{U}  &  =-\alpha\left(  \alpha-1\right)  U_{0}\left[  \frac{\pi^{2}}%
{4}\frac{a}{R}-3\frac{R}{a}-6\ln\left(  1+e^{-2R/a}\right)  \right.
\nonumber\\
&  \left.  -3\frac{a}{R}f_{2}\left(  e^{-2R/a}\right)  +3\frac{R}{a}\tanh
\frac{R}{a}\right]  .
\end{align}

\section{The effect of boundaries and potentials to ideal quantum gases}

In this section, we calculate the thermodynamic properties of ideal gases in
confined space with an external potential. First, we discuss the properties of
ideal gases under the condition of weak degeneration. We give the virial
expansion of the equation of state. Second, we discuss the properties of ideal
gases under the condition of complete degeneration. We obtain the asymptotic
expression of the specific heat at low temperatures and high densities.

\subsection{Weak degenerate ideal quantum gases in confined space with
potentials}

In this section, we consider weak degenerate ideal quantum gases. The virial
expression of the equation of state is given, which is modified by the
boundary and the potential.

According to the equation of state Eq. (\ref{EOS fermi/bose}), we obtain%
\begin{equation}
\frac{pV}{NkT}=\frac{\sum_{l=0,\frac{1}{2},1,\cdots}^{\infty}B_{l}\left[
\lambda^{2}/\left(  4\pi\right)  \right]  ^{l}h_{D/2+1-l}\left(  z\right)
}{\sum_{s=0,\frac{1}{2},1,\cdots}^{\infty}B_{s}\left[  \lambda^{2}/\left(
4\pi\right)  \right]  ^{s}h_{D/2-s}\left(  z\right)  }, \label{EOS 1}%
\end{equation}%
\begin{equation}
\frac{n\lambda^{D}}{g}=\sum_{l=0,\frac{1}{2},1,\cdots}^{\infty}\frac{B_{l}%
}{B_{0}}\left(  \frac{\lambda^{2}}{4\pi}\right)  ^{l}h_{D/2-l}\left(
z\right)  , \label{EOS 2}%
\end{equation}
where $n=N/V$ is the number density and $g$ is a weight factor arising form
the internal structure of the particle; noting that $B_{0}=V$, the volume.
Truncating Eqs. (\ref{EOS 1}) and (\ref{EOS 2}) up to $B_{1}$ and then
expanding them with respect to $z$, we obtain%
\begin{align}
\frac{pV}{NkT}  &  =\frac{B_{0}h_{D/2+1}\left(  z\right)  +\frac{\lambda
}{\sqrt{4\pi}}B_{1/2}h_{D/2+1/2}\left(  z\right)  +\frac{\lambda^{2}}{4\pi
}B_{1}h_{D/2}\left(  z\right)  }{B_{0}h_{D/2}\left(  z\right)  +\frac{\lambda
}{\sqrt{4\pi}}B_{1/2}h_{D/2-1/2}+\frac{\lambda^{2}}{4\pi}B_{1}h_{D/2-1}\left(
z\right)  }\nonumber\\
&  =1\mp\frac{2^{-\mathcal{D}/2}\left(  2\pi B_{0}+\sqrt{2\pi}\lambda
B_{1/2}+\lambda^{2}B_{1}\right)  }{4\pi B_{0}+2\sqrt{\pi}\lambda
B_{1/2}+\lambda^{2}B_{1}}z+\cdots, \label{EOS 3}%
\end{align}
and%
\begin{equation}
\frac{n\lambda^{D}}{g}=h_{D/2}\left(  z\right)  +\frac{\lambda}{\sqrt{4\pi}%
}\frac{B_{1/2}}{B_{0}}h_{D/2-1/2}\left(  z\right)  +\frac{\lambda^{2}}{4\pi
}\frac{B_{1}}{B_{0}}h_{D/2-1}\left(  z\right)  +\cdots. \label{EOS 4}%
\end{equation}
Inverting the series in Eq. (\ref{EOS 4}), we obtain an expression for $z$ in
powers of $n\lambda^{D}/g$,
\begin{align}
z  &  =\frac{1}{1+\lambda B_{1/2}/\left(  2\sqrt{\pi}B_{0}\right)
+\lambda^{2}B_{1}/\left(  4\pi B_{0}\right)  }\frac{n\lambda^{D}}%
{g}\nonumber\\
&  \mp\frac{2^{5-D/2}\pi^{2}B_{0}^{2}\left(  2\pi B_{0}+\sqrt{2\pi}\lambda
B_{1/2}+\lambda^{2}B_{1}\right)  }{\left(  4\pi B_{0}+2\sqrt{\pi}\lambda
B_{1/2}+\lambda^{2}B_{1}\right)  ^{3}}\left(  \frac{n\lambda^{D}}{g}\right)
^{2}+\cdots, \label{z}%
\end{align}
Substituting Eq. (\ref{z}) into Eq. (\ref{EOS 3}), the virial expansion is%
\begin{equation}
\frac{pV}{NkT}=1\mp\frac{n\lambda^{D}}{g}\left[  \frac{1}{2^{D/2+1}}%
-\frac{\sqrt{2}-1}{2^{\left(  D+3\right)  /2}}\frac{\lambda B_{1/2}}{\sqrt
{\pi}B_{0}}-\frac{2-\sqrt{2}}{2^{\left(  D+7\right)  /2}}\frac{\lambda
^{3}B_{1/2}B_{1}}{\pi^{3/2}B_{0}^{2}}\right]  +\cdots.
\end{equation}

The other thermodynamic quantities are the internal energy
\begin{align}
\frac{U}{NkT}  &  =\left(  \frac{D}{2}-\frac{\lambda B_{1/2}}{4\sqrt{\pi}%
B_{0}}-\frac{\lambda^{2}B_{1}}{4\pi B_{0}}+\frac{3\lambda^{3}B_{1/2}B_{1}%
}{16\pi^{3/2}B_{0}^{2}}\right) \nonumber\\
&  \mp\frac{n\lambda^{D}}{g}\left[  \frac{1}{2^{D/2+2}}-\frac{\sqrt{2}%
-1}{2^{\left(  D+5\right)  /2}}\left(  D+1\right)  \frac{\lambda B_{1/2}%
}{\sqrt{\pi}B_{0}}\right. \\
&  \left.  -\frac{2-\sqrt{2}}{2^{\left(  D+9\right)  /2}}\left(  D+3\right)
\frac{\lambda^{3}B_{1/2}B_{1}}{\pi^{3/2}B_{0}^{2}}\right]  +\cdots,
\end{align}
the Helmholtz free energy
\begin{align}
\frac{F}{NkT}  &  =\left[  \ln\left(  \frac{n\lambda^{D}}{g}\right)
-1-\frac{\lambda B_{1/2}}{2\sqrt{\pi}B_{0}}-\frac{\lambda^{2}B_{1}}{4\pi
B_{0}}+\frac{\lambda^{3}B_{1/2}B_{1}}{8\pi^{3/2}B_{0}^{2}}\right] \nonumber\\
&  \mp\frac{n\lambda^{D}}{g}\left[  \frac{1}{2^{D/2+1}}-\frac{\sqrt{2}%
-1}{2^{\left(  D+3\right)  /2}}\frac{\lambda B_{1/2}}{\sqrt{\pi}B_{0}}\right.
\nonumber\\
&  \left.  -\frac{\sqrt{2}-1}{2^{D/2+3}}\frac{\lambda^{3}B_{1/2}B_{1}}%
{\pi^{3/2}B_{0}^{2}}\right]  +\cdots,
\end{align}
the entropy
\begin{align}
\frac{S}{Nk}  &  =\left[  1+\frac{D}{2}+\frac{\lambda B_{1/2}}{4\sqrt{\pi
}B_{0}}+\frac{\lambda^{3}B_{1/2}B_{1}}{16\pi^{3/2}B_{0}^{2}}-\ln\left(
\frac{n\lambda^{D}}{g}\right)  \right] \nonumber\\
&  \mp\frac{n\lambda^{D}}{g}\left[  \frac{1}{2^{D/2+2}}\left(  D-2\right)
-\frac{\sqrt{2}-1}{2^{\left(  D+5\right)  /2}}\left(  D-1\right)
\frac{\lambda B_{1/2}}{\sqrt{\pi}B_{0}}\right. \nonumber\\
&  \left.  -\frac{\sqrt{2}-1}{2^{D/2+4}}\left(  D+1\right)  \frac{\lambda
^{3}B_{1/2}B_{1}}{\pi^{3/2}B_{0}{}^{2}}\right]  +\cdots,
\end{align}
and the specific heat
\begin{align}
\frac{C_{V}}{Nk}  &  =\left(  \frac{D}{2}-\frac{\lambda B_{1/2}}{8\sqrt{\pi
}B_{0}}-\frac{3\lambda^{3}B_{1/2}B_{1}}{32\pi^{3/2}B_{0}{}^{2}}\right)
\nonumber\\
&  \pm\frac{n\lambda^{D}}{g}\left[  \frac{1}{2^{D/2+3}}\left(  D^{2}%
-2D\right)  -\frac{\sqrt{2}-1}{2^{\left(  D+7\right)  /2}}\left(
D^{2}-1\right)  \frac{\lambda B_{1/2}}{\sqrt{\pi}B_{0}}\right. \nonumber\\
&  \left.  -\frac{\sqrt{2}-1}{2^{D/2+5}}\left(  D^{2}+4D+3\right)
\frac{\lambda^{3}B_{1/2}B_{1}}{\pi^{3/2}B_{0}^{2}}\right]  +\cdots.
\end{align}

\subsection{Completely degenerate ideal quantum gases in confined space with
potentials}

In this section, we discuss the property of completely degenerate ideal gases.
We consider ideal Bose gases in two dimensions and ideal Fermi gases in two
and three dimensions. We obtain asymptotic expressions of the chemical
potential and specific heat at low temperatures and high densities for Bose
and Fermi gases, which show the influence of the boundary and potential.

\subsubsection{Ideal Fermi gases in two-dimensional confined space with
potentials}

From Eqs. (\ref{Cv fermi/bose}) and (\ref{EOS fermi/bose}), for a Fermi gas,
when $D=2$, the specific heat and the number density are
\begin{align}
&  \frac{C_{V}}{Nk}\nonumber\\
&  =\frac{\sum_{l=0,\frac{1}{2},1,\cdots}^{\infty}B_{l}\left(  l^{2}%
-3l+2\right)  \left[  \lambda^{2}/\left(  4\pi\right)  \right]  ^{l}%
f_{2-l}\left(  z\right)  }{\sum_{s=0,\frac{1}{2},1,\cdots}^{\infty}%
B_{s}\left[  \lambda^{2}/\left(  4\pi\right)  \right]  ^{s}f_{1-s}\left(
z\right)  }\nonumber\\
&  -\frac{\left\{  \sum_{l=0,\frac{1}{2},1,\cdots}^{\infty}B_{l}\left(
1-l\right)  \left[  \lambda^{2}/\left(  4\pi\right)  \right]  ^{l}%
f_{1-l}\left(  z\right)  \right\}  ^{2}}{\sum_{s=0,\frac{1}{2},1,\cdots
}^{\infty}\sum_{j=0,\frac{1}{2},1,\cdots}^{\infty}B_{s}B_{j}\left[
\lambda^{2}/\left(  4\pi\right)  \right]  ^{s+j}f_{1-s}\left(  z\right)
f_{-j}\left(  z\right)  }, \label{Cv fermi D=2}%
\end{align}
and%
\begin{equation}
\frac{n\lambda^{2}}{g}=\sum_{l=0,\frac{1}{2},1,\cdots}^{\infty}\frac{B_{l}%
}{B_{0}}\left(  \frac{\lambda^{2}}{4\pi}\right)  ^{l}f_{1-l}\left(  z\right)
. \label{N fermi D=2}%
\end{equation}
Truncating Eqs. (\ref{Cv fermi D=2}) and (\ref{N fermi D=2}) up to $B_{1}$ and
then using \cite{pathria2011statistical}
\begin{align}
f_{\nu}\left(  z\right)   &  =\frac{\left(  \ln z\right)  ^{\nu}}%
{\Gamma\left(  \nu+1\right)  }\left[  1+\nu\left(  \nu-1\right)  \frac{\pi
^{2}}{6}\left(  \ln z\right)  ^{-2}\right. \nonumber\\
&  \left.  +\nu\left(  \nu-1\right)  \left(  \nu-2\right)  \left(
\nu-3\right)  \frac{7\pi^{4}}{360}\left(  \ln z\right)  ^{-4}+\cdots\right]  ,
\label{Sommerfeld}%
\end{align}
we obtain $C_{V}/\left(  Nk\right)  $ and $n\lambda^{2}/g$ in powers of
$kT/\mu$,
\begin{align}
\frac{C_{V}}{Nk}  &  =\frac{\pi^{2}}{3}\frac{kT}{\mu}-\frac{\pi\lambda
B_{1/2}}{6B_{0}}\left(  \frac{kT}{\mu}\right)  ^{3/2}+\left(  \frac
{\lambda^{2}B_{1/2}^{2}}{6B_{0}^{2}}-\frac{\pi\lambda^{2}B_{1}}{12B_{0}%
}\right)  \left(  \frac{kT}{\mu}\right)  ^{2}\nonumber\\
&  -\left(  \frac{\lambda^{3}B_{1/2}^{3}}{6\pi B_{0}^{3}}-\frac{\lambda
^{3}B_{1/2}B_{1}}{8B_{0}^{2}}\right)  \left(  \frac{kT}{\mu}\right)
^{5/2}+\cdots, \label{Cv mu}%
\end{align}
and
\begin{align}
\frac{n\lambda^{2}}{g}  &  =\frac{\mu}{kT}\left[  1+\frac{\lambda^{2}B_{1}%
}{4\pi B_{0}}\frac{kT}{\mu}+\frac{\lambda B_{1/2}}{\pi B_{0}}\left(  \frac
{kT}{\mu}\right)  ^{3/2}\right. \nonumber\\
&  \left.  -\frac{\pi\lambda B_{1/2}}{24B_{0}}\left(  \frac{kT}{\mu}\right)
^{5/2}-\frac{7\pi^{3}\lambda B_{1/2}}{384B_{0}}\left(  \frac{kT}{\mu}\right)
^{9/2}+\cdots\right]  . \label{n,mu}%
\end{align}
From Eq. (\ref{n,mu}), we have
\begin{align}
\mu &  =\varepsilon_{F}\left[  1-\frac{\lambda B_{1/2}}{\pi B_{0}}\left(
\frac{kT}{\mu}\right)  ^{1/2}+\left(  \frac{\lambda^{2}B_{1/2}^{2}}{\pi
^{2}B_{0}^{2}}-\frac{\lambda^{2}B_{1}}{4\pi B_{0}}\right)  \frac{kT}{\mu
}\right. \nonumber\\
&  \left.  -\left(  \frac{\lambda^{3}B_{1/2}^{3}}{\pi^{3}B_{0}^{3}}%
-\frac{\lambda^{3}B_{1/2}B_{1}}{2\pi^{2}B_{0}^{2}}\right)  \left(  \frac
{kT}{\mu}\right)  ^{3/2}\right. \nonumber\\
&  \left.  +\left(  \frac{\lambda^{4}B_{1/2}^{4}}{\pi^{4}B_{0}^{4}}%
-\frac{3\lambda^{4}B_{1/2}^{2}B_{1}}{4\pi^{3}B_{0}^{3}}+\frac{\lambda^{4}%
B_{1}^{2}}{16\pi^{2}B_{0}^{2}}\right)  \left(  \frac{kT}{\mu}\right)
^{2}+\cdots\right]  , \label{mu}%
\end{align}
where $\varepsilon_{F}=nh^{2}/\left(  2\pi gm\right)  $ is the two-dimensional
Fermi energy. Inverting the series in Eq. (\ref{mu}) to obtain an expansion
for $\mu$ in powers of $kT/\varepsilon_{F}$,
\begin{align}
\mu &  =\varepsilon_{F}\left[  1-\frac{\lambda B_{1/2}}{\pi B_{0}}\left(
\frac{kT}{\varepsilon_{F}}\right)  ^{1/2}+\left(  \frac{\lambda^{2}B_{1/2}%
^{2}}{2\pi^{2}B_{0}^{2}}-\frac{\lambda^{2}B_{1}}{4\pi B_{0}}\right)  \frac
{kT}{\varepsilon_{F}}\right. \nonumber\\
&  \left.  -\left(  \frac{\lambda^{3}B_{1/2}^{3}}{8\pi^{3}B_{0}^{3}}%
-\frac{\lambda^{3}B_{1/2}B_{1}}{8\pi^{2}B_{0}^{2}}\right)  \left(  \frac
{kT}{\varepsilon_{F}}\right)  ^{3/2}\right. \nonumber\\
&  \left.  +\left(  \frac{129\lambda^{5}B_{1/2}^{5}}{128\pi^{5}B_{0}^{5}%
}-\frac{65\lambda^{5}B_{1/2}^{3}B_{1}}{64\pi^{4}B_{0}^{4}}+\frac{25\lambda
^{5}B_{1/2}B_{1}^{2}}{128\pi^{3}B_{0}^{3}}\right)  \left(  \frac
{kT}{\varepsilon_{F}}\right)  ^{5/2}\right. \nonumber\\
&  \left.  +\left(  \frac{2\lambda^{6}B_{1/2}^{6}}{\pi^{6}B_{0}^{6}}%
-\frac{7\lambda^{6}B_{1/2}^{4}B_{1}}{4\pi^{5}B_{0}^{5}}+\frac{3\lambda
^{6}B_{1/2}^{2}B_{1}^{2}}{16\pi^{4}B_{0}^{4}}+\frac{\lambda^{6}B_{1}^{3}%
}{64\pi^{3}B_{0}^{3}}\right)  \left(  \frac{kT}{\varepsilon_{F}}\right)
^{3}+\cdots\right]  , \label{mu series}%
\end{align}
and then substituting Eq. (\ref{mu series}) into Eq. (\ref{Cv mu}), we obtain
the asymptotic expression of the specific heat at low temperatures and high
densities
\begin{align}
\frac{C_{V}}{Nk}  &  =\frac{\pi^{2}}{3}\frac{kT}{\varepsilon_{F}}+\frac
{\pi\lambda B_{1/2}}{6B_{0}}\left(  \frac{kT}{\varepsilon_{F}}\right)
^{3/2}+\frac{\lambda^{2}B_{1/2}^{2}}{12B_{0}^{2}}\left(  \frac{kT}%
{\varepsilon_{F}}\right)  ^{2}\nonumber\\
&  +\left(  \frac{3\lambda^{3}B_{1/2}^{3}}{16\pi B_{0}^{3}}-\frac{5\lambda
^{3}B_{1/2}B_{1}}{48B_{0}^{2}}\right)  \left(  \frac{kT}{\varepsilon_{F}%
}\right)  ^{5/2}\nonumber\\
&  +\left(  \frac{\lambda^{4}B_{1/2}^{4}}{4\pi^{2}B_{0}^{4}}-\frac{\lambda
^{4}B_{1/2}^{2}B_{1}}{8\pi B_{0}^{3}}-\frac{\lambda^{4}B_{1}^{2}}{48B_{0}^{2}%
}\right)  \left(  \frac{kT}{\varepsilon_{F}}\right)  ^{3}+\cdots.
\label{Cv 2D fermi}%
\end{align}

From Eq. (\ref{B0 B1/2 B1}),%
\begin{equation}
\frac{\Delta B_{1}}{B_{0}}=\frac{B_{1}-B_{1}^{\left(  0\right)  }}%
{B_{0}^{\left(  0\right)  }}=-\frac{2m\tilde{U}}{\hbar^{2}},
\end{equation}
the effect of the boundary and the potential to the specific heat is
\begin{equation}
\frac{\Delta C_{V}}{Nk}=\left(  -\frac{\pi^{2}\tilde{U}^{2}}{3\varepsilon
_{F}^{2}}+\frac{5\pi^{3/2}\tilde{U}\sqrt{m\varepsilon_{F}}\hbar B_{1/2}%
}{6\sqrt{2}m\varepsilon_{F}^{2}B_{0}}+\frac{\pi\tilde{U}\hbar^{2}B_{1/2}^{2}%
}{m\varepsilon_{F}^{2}B_{0}^{2}}+\frac{\pi^{2}\tilde{U}\hbar^{2}B_{1}%
}{3m\varepsilon_{F}^{2}B_{0}}\right)  \frac{kT}{\varepsilon_{F}}+\cdots,
\end{equation}
where $\Delta C_{V}=C_{V}-C_{V}^{\left(  0\right)  }$. When the volume
$V\rightarrow\infty$, the boundary effect vanishes and the external potential
effect is%
\begin{equation}
\left(  \frac{\Delta C_{V}}{Nk}\right)  _{V\rightarrow\infty}=-\frac{\pi
^{2}\tilde{U}^{2}}{3\varepsilon_{F}^{2}}\frac{kT}{\varepsilon_{F}}+\cdots.
\label{dCv 2D fermi}%
\end{equation}

Now, we consider the effect of external potentials. From Eqs.
(\ref{Cv 2D fermi}) and (\ref{dCv 2D fermi}), we obtain
\begin{equation}
\left\vert \frac{\Delta C_{V}}{C_{V}}\right\vert \sim\frac{\tilde{U}^{2}%
}{\varepsilon_{F}^{2}}.
\end{equation}
It is clear that the effect of the external potentials is independent of the
temperature. The Fermi energy of Cu, is approximately $4.6%
\operatorname{eV}%
$, and if we take the external potential given by Eq. (\ref{Gauss}),
$\tilde{U}$ is determined by Eq. (\ref{Gauss Utilde}). Let $a/R\sim0.3$ and
$U_{0}=2%
\operatorname{eV}%
$, $\tilde{U}$ is approximately $-1.6%
\operatorname{eV}%
$, the effect of the external potential is approximate
\begin{equation}
\left\vert \frac{\Delta C_{V}}{C_{V}}\right\vert \sim0.12.
\end{equation}

Performing the same procedure, we obtain the equation of state and the other
thermodynamic quantities: the equation of state,
\begin{align*}
\frac{pV}{NkT}  &  =\frac{2}{5}\frac{\varepsilon_{F}}{kT}\left[
1-\frac{2\lambda B_{1/2}}{3\pi B_{0}}\left(  \frac{kT}{\varepsilon_{F}%
}\right)  ^{1/2}+\left(  \frac{\lambda^{3}B_{1/2}^{3}}{4\pi^{3}B_{0}^{3}%
}-\frac{\lambda^{3}B_{1/2}B_{1}}{4\pi^{2}B_{0}^{2}}\right)  \left(  \frac
{kT}{\varepsilon_{F}}\right)  ^{3/2}\right. \\
&  \left.  +\left(  \frac{\pi^{2}}{3}-\frac{\lambda^{4}B_{1/2}^{4}}{6\pi
^{4}B_{0}^{4}}+\frac{\lambda^{4}B_{1/2}^{2}B_{1}}{4\pi^{3}B_{0}^{3}}%
-\frac{\lambda^{4}B_{1}^{2}}{16\pi^{2}B_{0}^{2}}\right)  \left(  \frac
{kT}{\varepsilon_{F}}\right)  ^{2}+\cdots\right]  ,
\end{align*}
the internal energy,
\begin{align}
\frac{U}{N}  &  =\frac{1}{2}\varepsilon_{F}\left[  1-\frac{4\lambda B_{1/2}%
}{3\pi B_{0}}\left(  \frac{kT}{\varepsilon_{F}}\right)  ^{1/2}+\left(
\frac{\lambda^{2}B_{1/2}^{2}}{\pi^{2}B_{0}^{2}}-\frac{\lambda^{2}B_{1}}{2\pi
B_{0}}\right)  \frac{kT}{\varepsilon_{F}}\right. \nonumber\\
&  \left.  -\left(  \frac{\lambda^{3}B_{1/2}^{3}}{2\pi^{3}B_{0}^{3}}%
-\frac{\lambda^{3}B_{1/2}B_{1}}{2\pi^{2}B_{0}^{2}}\right)  \left(  \frac
{kT}{\varepsilon_{F}}\right)  ^{3/2}\right. \nonumber\\
&  \left.  +\left(  \frac{\pi^{2}}{3}+\frac{\lambda^{4}B_{1/2}^{4}}{6\pi
^{4}B_{0}^{4}}-\frac{\lambda^{4}B_{1/2}^{2}B_{1}}{4\pi^{3}B_{0}^{3}}%
+\frac{\lambda^{4}B_{1}^{2}}{16\pi^{2}B_{0}^{2}}\right)  \left(  \frac
{kT}{\varepsilon_{F}}\right)  ^{2}+\cdots\right]  ,
\end{align}
the Helmholtz free energy,
\begin{align}
\frac{F}{N}  &  =\frac{1}{2}\varepsilon_{F}\left[  1-\frac{4\lambda B_{1/2}%
}{3\pi B_{0}}\left(  \frac{kT}{\varepsilon_{F}}\right)  ^{1/2}+\left(
\frac{\lambda^{2}B_{1/2}^{2}}{\pi^{2}B_{0}^{2}}-\frac{\lambda^{2}B_{1}}{2\pi
B_{0}}\right)  \frac{kT}{\varepsilon_{F}}\right. \nonumber\\
&  \left.  -\left(  \frac{\lambda^{3}B_{1/2}^{3}}{2\pi^{3}B_{0}^{3}}%
-\frac{\lambda^{3}B_{1/2}B_{1}}{2\pi^{2}B_{0}^{2}}\right)  \left(  \frac
{kT}{\varepsilon_{F}}\right)  ^{3/2}\right. \nonumber\\
&  \left.  +\left(  -\frac{\pi^{2}}{3}+\frac{\lambda^{4}B_{1/2}^{4}}{6\pi
^{4}B_{0}^{4}}-\frac{\lambda^{4}B_{1/2}^{2}B_{1}}{4\pi^{3}B_{0}^{3}}%
+\frac{\lambda^{4}B_{1}^{2}}{16\pi^{2}B_{0}^{2}}\right)  \left(  \frac
{kT}{\varepsilon_{F}}\right)  ^{2}+\cdots\right]  ,
\end{align}
and the entropy,
\begin{align}
\frac{S}{Nk}  &  =\frac{\pi^{2}}{3}\frac{kT}{\varepsilon_{F}}+\frac{\pi\lambda
B_{1/2}}{6B_{0}}\left(  \frac{kT}{\varepsilon_{F}}\right)  ^{3/2}%
+\frac{\lambda^{2}B_{1/2}^{2}}{12B_{0}^{2}}\left(  \frac{kT}{\varepsilon_{F}%
}\right)  ^{2}\nonumber\\
&  +\left(  \frac{3\lambda^{3}B_{1/2}^{3}}{16\pi B_{0}^{3}}-\frac{5\lambda
^{3}B_{1/2}B_{1}}{48B_{0}^{2}}\right)  \left(  \frac{kT}{\varepsilon_{F}%
}\right)  ^{5/2}\nonumber\\
&  -\left(  \frac{\lambda^{4}B_{1/2}^{4}}{4\pi^{2}B_{0}^{4}}-\frac{\lambda
^{4}B_{1/2}^{2}B_{1}}{8\pi B_{0}^{3}}-\frac{\lambda^{4}B_{1}^{2}}{48B_{0}^{2}%
}\right)  \left(  \frac{kT}{\varepsilon_{F}}\right)  ^{3}+\cdots.
\end{align}

\subsubsection{Ideal Fermi gases in three-dimensional confined space with
potentials}

For a Fermi gas in three-dimensional confined space, the specific heat and the
number density are
\begin{align}
&  \frac{C_{V}}{Nk}\nonumber\\
&  =\frac{\sum_{l=0,\frac{1}{2},1,\cdots}^{\infty}B_{l}\left(  l^{2}%
-4l+\frac{15}{4}\right)  \left[  \lambda^{2}/\left(  4\pi\right)  \right]
^{l}f_{5/2-l}\left(  z\right)  }{\sum_{s=0,\frac{1}{2},1,\cdots}^{\infty}%
B_{s}\left[  \lambda^{2}/\left(  4\pi\right)  \right]  ^{s}f_{3/2-s}\left(
z\right)  }\nonumber\\
&  -\frac{\left\{  \sum_{l=0,\frac{1}{2},1,\cdots}^{\infty}B_{l}\left(
\frac{3}{2}-l\right)  \left[  \lambda^{2}/\left(  4\pi\right)  \right]
^{l}f_{3/2-l}\left(  z\right)  \right\}  ^{2}}{\sum_{s=0,\frac{1}{2},1,\cdots
}^{\infty}\sum_{j=0,\frac{1}{2},1,\cdots}^{\infty}B_{s}B_{j}\left[
\lambda^{2}/\left(  4\pi\right)  \right]  ^{s+j}f_{3/2-s}\left(  z\right)
f_{1/2-j}\left(  z\right)  } \label{Cv fermi 3D}%
\end{align}
and
\begin{equation}
\frac{n\lambda^{3}}{g}=\sum_{l=0,\frac{1}{2},1,\cdots}^{\infty}\frac{B_{l}%
}{B_{0}}\left(  \frac{\lambda^{2}}{4\pi}\right)  ^{l}f_{3/2-l}\left(
z\right)  . \label{fermi 3D Number}%
\end{equation}

Truncating Eq. (\ref{Cv fermi 3D}) up to $B_{1}$ and then using Eq.
(\ref{Sommerfeld}), we achieve%
\begin{equation}
\frac{C_{V}}{Nk}=\frac{\pi^{2}}{2}\frac{kT}{\mu}-\frac{\pi^{2}\lambda B_{1/2}%
}{16B_{0}}\left(  \frac{kT}{\mu}\right)  ^{3/2}+\left(  \frac{3\pi^{2}%
B_{1/2}^{2}}{128B_{0}^{2}}-\frac{\pi\lambda^{2}B_{1}}{8B_{0}}\right)  \left(
\frac{kT}{\mu}\right)  ^{2}+\cdots\label{Cv mu 3D}%
\end{equation}
and
\begin{align}
\mu &  =\varepsilon_{F}\left[  1-\frac{\lambda B_{1/2}}{4B_{0}}\left(
\frac{kT}{\mu}\right)  ^{1/2}-\left(  \frac{\lambda^{2}B_{1}}{4\pi B_{0}%
}-\frac{5\lambda^{2}B_{1/2}^{2}}{64B_{0}^{2}}\right)  \frac{kT}{\mu}\right.
\nonumber\\
&  \left.  -\left(  \frac{5\lambda^{3}B_{1/2}^{3}}{192B_{0}^{3}}%
-\frac{5\lambda^{3}B_{1/2}B_{1}}{32\pi B_{0}^{2}}\right)  \left(  \frac
{kT}{\mu}\right)  ^{3/2}\right. \nonumber\\
&  \left.  -\left(  \frac{\pi^{2}}{12}-\frac{55\lambda^{4}B_{1/2}^{4}%
}{6144B_{0}^{4}}+\frac{5\lambda^{4}B_{1/2}^{2}B_{1}}{64\pi B_{0}^{3}}%
-\frac{5\lambda^{4}B_{1}^{2}}{64\pi^{2}B_{0}^{2}}\right)  \left(  \frac
{kT}{\mu}\right)  ^{2}-\cdots\right]  , \label{mu 3D}%
\end{align}
where $\varepsilon_{F}=\left(  h^{2}/2m\right)  \left[  3n/\left(  4\pi
g\right)  \right]  ^{2/3}$ is the three-dimensional Fermi energy. Inverting
the series in Eq. (\ref{mu 3D}), we obtain an expansion for $\mu$ in powers of
$kT/\varepsilon_{F}$,
\begin{align}
\mu &  =\varepsilon_{F}\left[  1-\frac{\lambda B_{1/2}}{4B_{0}}\left(
\frac{kT}{\varepsilon_{F}}\right)  ^{1/2}-\left(  \frac{\lambda^{2}B_{1}}{4\pi
B_{0}}-\frac{3\lambda^{2}B_{1/2}^{2}}{64B_{0}^{2}}\right)  \frac
{kT}{\varepsilon_{F}}\right. \nonumber\\
&  \left.  -\left(  \frac{5\lambda^{3}B_{1/2}^{3}}{768B_{0}^{3}}-\frac
{\lambda^{3}B_{1/2}B_{1}}{16\pi B_{0}^{2}}\right)  \left(  \frac
{kT}{\varepsilon_{F}}\right)  ^{3/2}\right. \nonumber\\
&  \left.  -\left(  \frac{\pi^{2}}{12}-\frac{7\lambda^{4}B_{1/2}^{4}%
}{12288B_{0}^{4}}+\frac{\lambda^{4}B_{1/2}^{2}B_{1}}{128\pi B_{0}^{3}}%
-\frac{\lambda^{4}B_{1}^{2}}{64\pi^{2}B_{0}^{2}}\right)  \left(  \frac
{kT}{\varepsilon_{F}}\right)  ^{2}+\cdots\right]  . \label{mu series 3D}%
\end{align}
Substituting Eq. (\ref{mu series 3D}) into Eq. (\ref{Cv mu 3D}), we obtain the
asymptotic expression of specific heat at low temperatures and high densities
\begin{align}
\frac{C_{V}}{Nk}  &  =\frac{\pi^{2}}{2}\frac{kT}{\varepsilon_{F}}+\frac
{\pi^{2}\lambda B_{1/2}}{16B_{0}}\left(  \frac{kT}{\varepsilon_{F}}\right)
^{3/2}+\frac{\pi^{2}\lambda^{2}B_{1/2}^{2}}{128B_{0}^{2}}\left(  \frac
{kT}{\varepsilon_{F}}\right)  ^{2}\nonumber\\
&  +\left(  \frac{25\pi^{2}\lambda^{3}B_{1/2}^{3}}{3072B_{0}^{3}}-\frac
{7\pi\lambda^{3}B_{1/2}B_{1}}{128B_{0}^{2}}\right)  \left(  \frac
{kT}{\varepsilon_{F}}\right)  ^{5/2}\nonumber\\
&  +\left(  \frac{\pi^{4}}{24}+\frac{9\pi^{2}\lambda^{4}B_{1/2}^{4}}%
{4096B_{0}^{4}}-\frac{9\pi\lambda^{4}B_{1/2}^{2}B_{1}}{1024B_{0}^{3}}%
-\frac{5\lambda^{4}B_{1}^{2}}{128B_{0}^{2}}\right)  \left(  \frac
{kT}{\varepsilon_{F}}\right)  ^{3}+\cdots. \label{Cv 3D fermi}%
\end{align}

Using
\[
B_{0}=B_{0}^{\left(  0\right)  },\ B_{1/2}=B_{1/2}^{\left(  0\right)
},\ \text{and }\frac{\Delta B_{1}}{B_{0}}=\frac{B_{1}-B_{1}^{\left(  0\right)
}}{B_{0}^{\left(  0\right)  }}=-\frac{2m\tilde{U}}{\hbar^{2}},
\]
we can see that the effect of boundary and potential to the specific heat is
\begin{equation}
\frac{\Delta C_{V}}{Nk}=\left(  -\frac{5\pi^{2}\tilde{U}^{2}}{8\varepsilon
_{F}^{2}}+\frac{7\pi^{5/2}\tilde{U}\sqrt{m\varepsilon_{F}}\hbar B_{1/2}%
}{16\sqrt{2}m\varepsilon_{F}^{2}B_{0}}+\frac{9\pi^{3}\tilde{U}\hbar^{2}%
B_{1/2}^{2}}{128m\varepsilon_{F}^{2}B_{0}^{2}}+\frac{5\pi^{2}\tilde{U}%
\hbar^{2}B_{1}}{8m\varepsilon_{F}^{2}B_{0}}\right)  \frac{kT}{\varepsilon_{F}%
}+\cdots.
\end{equation}
When the volume $V\rightarrow\infty$, the boundary effect vanishes and the
effect of the external potential is
\begin{equation}
\left(  \frac{\Delta C_{V}}{Nk}\right)  _{V\rightarrow\infty}=-\frac{5\pi
^{2}\tilde{U}^{2}}{8\varepsilon_{F}^{2}}\frac{kT}{\varepsilon_{F}}+\cdots.
\label{dCv 3D fermi}%
\end{equation}

Now, we consider the effect of external potentials. From Eqs.
(\ref{Cv 3D fermi}) and (\ref{dCv 3D fermi}), we obtain
\begin{equation}
\left\vert \frac{\Delta C_{V}}{C_{V}}\right\vert \sim\frac{5}{4}\frac
{\tilde{U}^{2}}{\varepsilon_{F}^{2}}.
\end{equation}
It is clear that the effect of the external potentials is independent of the
temperature. The Fermi energy of electronic gases in metal is from $1.5$ to
$15%
\operatorname{eV}%
$. The Fermi energy of Cu, for instance, is approximately $7%
\operatorname{eV}%
$. If we take the external potential given by Eq. (\ref{Gauss}), $\tilde{U}$
is determined by Eq. (\ref{Gauss Utilde}). Let $a/R\sim0.3$ and $U_{0}=2%
\operatorname{eV}%
$, $\tilde{U}$ is approximately $-0.89%
\operatorname{eV}%
$, the effect of the external potential is approximately
\begin{equation}
\left\vert \frac{\Delta C_{V}}{C_{V}}\right\vert \sim0.02.
\end{equation}

Performing the same procedure, we obtain the equation of state and the other
thermodynamic quantities: the equation of state,
\begin{align*}
\frac{pV}{NkT}  &  =\frac{2}{5}\frac{\varepsilon_{F}}{kT}\left[
1-\frac{5\lambda B_{1/2}}{32B_{0}}\left(  \frac{kT}{\varepsilon_{F}}\right)
^{1/2}+\left(  \frac{25\lambda^{3}B_{1/2}^{3}}{3072B_{0}^{3}}-\frac
{5\lambda^{3}B_{1/2}B_{1}}{64\pi B_{0}^{2}}\right)  \left(  \frac
{kT}{\varepsilon_{F}}\right)  ^{3/2}\right. \\
&  \left.  +\left(  \frac{5\pi^{2}}{12}-\frac{35\lambda^{4}B_{1/2}^{4}%
}{12288B_{0}^{4}}+\frac{5\lambda^{4}B_{1/2}^{2}B_{1}}{128\pi B_{0}^{3}}%
-\frac{5\lambda^{4}B_{1}^{2}}{64\pi^{2}B_{0}^{2}}\right)  \left(  \frac
{kT}{\varepsilon_{F}}\right)  ^{2}+\cdots\right]  ,
\end{align*}
the internal energy,
\begin{align}
\frac{U}{N}  &  =\frac{3}{5}\varepsilon_{F}\left[  1-\frac{5\lambda B_{1/2}%
}{16B_{0}}\left(  \frac{kT}{\varepsilon_{F}}\right)  ^{1/2}+\left(
\frac{5\lambda^{2}B_{1/2}^{2}}{64B_{0}^{2}}-\frac{5\lambda^{2}B_{1}}{12\pi
B_{0}}\right)  \frac{kT}{\varepsilon_{F}}\right. \nonumber\\
&  \left.  +\left(  \frac{5\lambda^{3}B_{1/2}B_{1}}{32\pi B_{0}^{2}}%
-\frac{25\lambda^{3}B_{1/2}^{3}}{1536B_{0}^{3}}\right)  \left(  \frac
{kT}{\varepsilon_{F}}\right)  ^{3/2}\right. \nonumber\\
&  \left.  +\left(  \frac{5\pi^{2}}{12}+\frac{35\lambda^{4}B_{1/2}^{4}%
}{12288B_{0}^{4}}-\frac{5\lambda^{4}B_{1/2}^{2}B_{1}}{128\pi B_{0}^{3}}%
+\frac{5\lambda^{4}B_{1}^{2}}{64\pi^{2}B_{0}^{2}}\right)  \left(  \frac
{kT}{\varepsilon_{F}}\right)  ^{2}+\cdots\right]  ,
\end{align}
the Helmholtz free energy,
\begin{align}
\frac{F}{N}  &  =\frac{3}{5}\varepsilon_{F}\left[  1-\frac{5\lambda B_{1/2}%
}{16B_{0}}\left(  \frac{kT}{\varepsilon_{F}}\right)  ^{1/2}+\left(
\frac{5\lambda^{2}B_{1/2}^{2}}{64B_{0}^{2}}-\frac{5\lambda^{2}B_{1}}{12\pi
B_{0}}\right)  \frac{kT}{\varepsilon_{F}}\right. \nonumber\\
&  \left.  +\left(  \frac{5\lambda^{3}B_{1/2}B_{1}}{32\pi B_{0}^{2}}%
-\frac{25\lambda^{3}B_{1/2}^{3}}{1536B_{0}^{3}}\right)  \left(  \frac
{kT}{\varepsilon_{F}}\right)  ^{3/2}\right. \nonumber\\
&  \left.  +\left(  -\frac{5\pi^{2}}{12}+\frac{35\lambda^{4}B_{1/2}^{4}%
}{12288B_{0}^{4}}-\frac{5\lambda^{4}B_{1/2}^{2}B_{1}}{128\pi B_{0}^{3}}%
+\frac{5\lambda^{4}B_{1}^{2}}{64\pi^{2}B_{0}^{2}}\right)  \left(  \frac
{kT}{\varepsilon_{F}}\right)  ^{2}+\cdots\right]  ,
\end{align}
and the entropy,
\begin{align}
\frac{S}{Nk}  &  =\frac{\pi^{2}}{2}\frac{kT}{\varepsilon_{F}}+\frac{\pi
^{2}\lambda B_{1/2}}{16B_{0}}\left(  \frac{kT}{\varepsilon_{F}}\right)
^{3/2}+\frac{\pi^{2}\lambda^{2}B_{1/2}^{2}}{128B_{0}^{2}}\left(  \frac
{kT}{\varepsilon_{F}}\right)  ^{2}\nonumber\\
&  +\left(  \frac{25\pi^{2}\lambda^{3}B_{1/2}^{3}}{3072B_{0}^{3}}-\frac
{7\pi\lambda^{3}B_{1/2}B_{1}}{128B_{0}^{2}}\right)  \left(  \frac
{kT}{\varepsilon_{F}}\right)  ^{5/2}\nonumber\\
&  +\left(  \frac{\pi^{4}}{24}+\frac{9\pi^{2}\lambda^{4}B_{1/2}^{4}}%
{4096B_{0}^{4}}-\frac{9\pi\lambda^{4}B_{1/2}^{2}B_{1}}{1024B_{0}^{3}}%
-\frac{5\lambda^{4}B_{1}^{2}}{128B_{0}^{2}}\right)  \left(  \frac
{kT}{\varepsilon_{F}}\right)  ^{3}+\cdots.
\end{align}

\subsubsection{Ideal Bose gases in two-dimensional confined space with
potentials}

For a Bose gas in two-dimensional confined space, the specific heat and the
number density are
\begin{align}
&  \frac{C_{V}}{Nk}=\frac{\sum_{l=0,\frac{1}{2},1,\cdots}^{\infty}B_{l}\left(
l^{2}-3l+2\right)  \left[  \lambda^{2}/\left(  4\pi\right)  \right]
^{l}g_{2-l}\left(  z\right)  }{\sum_{s=0,\frac{1}{2},1,\cdots}^{\infty}%
B_{s}\left[  \lambda^{2}/\left(  4\pi\right)  \right]  ^{s}g_{1-s}\left(
z\right)  }\nonumber\\
&  -\frac{\left\{  \sum_{l=0,\frac{1}{2},1,\cdots}^{\infty}B_{l}\left(
1-l\right)  \left[  \lambda^{2}/\left(  4\pi\right)  \right]  ^{l}%
g_{1-l}\left(  z\right)  \right\}  ^{2}}{\sum_{s=0,\frac{1}{2},1,\cdots
}^{\infty}\sum_{j=0,\frac{1}{2},1,\cdots}^{\infty}B_{s}B_{j}\left[
\lambda^{2}/\left(  4\pi\right)  \right]  ^{s+j}g_{-j}\left(  z\right)
g_{1-s}\left(  z\right)  }, \label{bose Cv/Nk 2D}%
\end{align}
and%

\begin{equation}
N=g\frac{V}{\lambda^{2}}\sum_{l=0,\frac{1}{2},1,\cdots}^{\infty}\frac{B_{l}%
}{B_{0}}\left(  \frac{\lambda^{2}}{4\pi}\right)  ^{l}g_{1-l}\left(  z\right)
. \label{bose N 2D}%
\end{equation}
Truncating Eq. (\ref{bose N 2D}) up to $B_{0}$ gives
\begin{equation}
\frac{n\lambda^{2}}{g}=-\ln\left(  1-z\right)  +\cdots,
\end{equation}
and then we achieve%
\begin{equation}
z=1-e^{-n\lambda^{2}/g}+\cdots.
\end{equation}
Truncating Eq. (\ref{bose Cv/Nk 2D}) up to $B_{1}$ and then using%
\begin{equation}
g_{1}\left(  z\right)  =-\ln\left(  1-z\right)
\end{equation}
and $\nu\not =1$,
\begin{align}
g_{\nu}\left(  z\right)   &  =\left(  1-z\right)  ^{\nu-1}\left[  -\nu
\Gamma\left(  -\nu\right)  -\frac{1}{2}\Gamma\left(  2-\nu\right)  \left(
1-z\right)  \right. \nonumber\\
&  \left.  -\frac{1}{24}\left(  3\nu+2\right)  \Gamma\left(  2-\nu\right)
\left(  1-z\right)  ^{2}+\cdots\right]  ,
\end{align}
where $\zeta\left(  \nu\right)  $ is the Riemann zeta function, we obtain%
\begin{align}
\frac{C_{V}}{k}  &  =\frac{\pi^{2}}{3}\frac{B_{0}}{\lambda^{2}}+\frac
{3\zeta\left(  \frac{3}{2}\right)  }{8\sqrt{\pi}}\frac{B_{1/2}}{\lambda
}+e^{-n\lambda^{2}/\left(  2g\right)  }\frac{3}{4}\frac{B_{1/2}}{\lambda
}\nonumber\\
&  -e^{-n\lambda^{2}/g}\left[  2\left(  1+\frac{n\lambda^{2}}{g}\right)
\frac{B_{0}}{\lambda^{2}}+\frac{3\zeta\left(  \frac{1}{2}\right)  }{8\sqrt
{\pi}}\frac{B_{1/2}}{\lambda}+\frac{\pi}{4}\frac{B_{1/2}^{2}}{\lambda^{2}%
B_{1}}\right]  +\cdots.
\end{align}

Moreover, the equation of state and the other thermodynamic quantities are the
equation of state,
\begin{align}
\frac{pV}{kT}  &  =\left[  \frac{\pi^{2}}{6}\frac{B_{0}}{\lambda^{2}}%
+\frac{\zeta\left(  \frac{3}{2}\right)  }{2\sqrt{\pi}}\frac{B_{1/2}}{\lambda
}+\frac{n\lambda^{2}}{g}\frac{B_{1}}{4\pi}\right] \nonumber\\
&  +e^{-n\lambda^{2}/\left(  2g\right)  }\frac{B_{1/2}}{\lambda}%
-e^{-n\lambda^{2}/g}\left[  \left(  1+\frac{\lambda^{2}n}{g}\right)
\frac{B_{0}}{\lambda^{2}}+\frac{\zeta\left(  \frac{1}{2}\right)  }{2\sqrt{\pi
}}\frac{B_{1/2}}{\lambda}\right]  +\cdots,
\end{align}
the internal energy,
\begin{align}
\frac{U}{kT}  &  =\left[  \frac{\pi^{2}}{6}\frac{B_{0}}{\lambda^{2}}%
+\frac{\zeta\left(  \frac{3}{2}\right)  }{4\sqrt{\pi}}\frac{B_{1/2}}{\lambda
}+\frac{n\lambda^{2}}{g}\frac{B_{1}}{4\pi}\right]  +e^{-n\lambda^{2}/\left(
2g\right)  }\frac{1}{2}\frac{B_{1/2}}{\lambda}\nonumber\\
&  -e^{-n\lambda^{2}/g}\left[  \left(  1+\frac{n\lambda^{2}}{g}\right)
\frac{B_{0}}{\lambda^{2}}+\frac{\zeta\left(  \frac{1}{2}\right)  }{4\sqrt{\pi
}}\frac{B_{1/2}}{\lambda}\right]  +\cdots,
\end{align}
the Helmholtz free energy,
\begin{align}
\frac{F}{kT}  &  =-\frac{\pi^{2}}{6}\frac{B_{0}}{\lambda^{2}}-\frac
{\zeta\left(  \frac{3}{2}\right)  }{2\sqrt{\pi}}\frac{B_{1/2}}{\lambda
}-\left(  1+\frac{n\lambda^{2}}{g}\right)  \frac{B_{1}}{4\pi}\nonumber\\
&  -e^{-n\lambda^{2}/\left(  2g\right)  }\frac{1}{2}\frac{B_{1/2}}{\lambda
}+e^{-n\lambda^{2}/g}\left(  \frac{B_{0}}{\lambda^{2}}+\frac{1}{2}\frac{B_{1}%
}{4\pi}\right)  +\cdots,
\end{align}
and the entropy,
\begin{align}
\frac{S}{k}  &  =\frac{\pi^{2}}{3}\frac{B_{0}}{\lambda^{2}}+\frac
{3\zeta\left(  \frac{3}{2}\right)  }{4\sqrt{\pi}}\frac{B_{1/2}}{\lambda
}\nonumber\\
&  +\left(  1+\frac{n\lambda^{2}}{g}\right)  \frac{B_{1}}{4\pi}+e^{-n\lambda
^{2}/\left(  2g\right)  }\frac{B_{1/2}}{\lambda}\nonumber\\
&  -e^{-n\lambda^{2}/g}\left[  \left(  2+\frac{n\lambda^{2}}{g}\right)
\frac{B_{0}}{\lambda^{2}}+\frac{\zeta\left(  \frac{1}{2}\right)  }{4\sqrt{\pi
}}\frac{B_{1/2}}{\lambda}+\frac{1}{2}\frac{B_{1}}{4\pi}\right]  +\cdots.
\end{align}

\section{Conclusion}

The heat kernel technique has been developed in mathematics and physics for
many years. In this paper, we employ the heat kernel technique to calculate
partition functions, grand potentials, and thermodynamic quantities of ideal
quantum gases in confined space with external potentials. Since the effect of
boundary and the external potential is reflected in the heat kernel
coefficient, we calculate the modification of the global heat kernel which
caused by potentials in confined space. At last, we consider the behaviors of
ideal quantum gases. Especially, by use of an analytic continuation, we
consider the application of the heat kernel technique to Fermi gases in which
the expansion will diverge when the fugacity $z>1$. We achieve the virial
expression under the condition of weak degeneration and the effects of the
boundary and the potential to thermodynamic quantities under the condition of
complete degeneration to ideal quantum gases.

The global heat kernel of the operator $-\Delta$ in confined space and the
$-\Delta+\frac{2m}{\hbar^{2}}V$ in free space have been discussed for many
years. Nevertheless, the method provided in this paper, in fact, is a way to
achieve the approximate global heat kernel of the operator $-\Delta+\frac
{2m}{\hbar^{2}}V$ in a confined space.

The method developed in the present paper can be used to calculate the heat
kernel for potentials. When the heat kernel is obtained, one can obtain
scattering phase shifts directly \cite{pang2012relation,li2015heat}. The
scattering phase shifts is the most important quantity in scattering theory
\cite{liu2014scattering,li2016scattering,gou2016covariant}. Therefore, the
method can also be applied to problems beyond statistical mechanics.

\acknowledgments

This work is supported in part by NSF of China under Grant No. 11575125.






\end{document}